\documentclass[superscriptaddress,nofootinbib]{revtex4}
\usepackage[dvips]{graphicx}
\usepackage{latexsym}
\usepackage{graphicx}
\usepackage[export]{adjustbox}
\usepackage{float}
\usepackage[utf8]{inputenc}
\usepackage[english]{babel}
\usepackage[intlimits]{amsmath}
\usepackage{mathtools}
\usepackage{amssymb}
\usepackage{verbatim}
\usepackage{amsfonts}
\usepackage{hyperref}
\usepackage{color}
  \usepackage{verbatim}

  \newcommand{\red}[1]{{\color{red} #1}}

  \definecolor{blue}{rgb}{0,0,1}
  \definecolor{green}{rgb}{0,.6,0}
  \definecolor{red}{rgb}{1,0,0}
  \definecolor{vio}{rgb}{1,0,1}
  \definecolor{uv}{rgb}{0.5,0,0.5}
  \definecolor{ama}{rgb}{0.3,0.3,0.3}
  \newcommand{\dif}{\mathrm{d}}
  \newcommand{\R}{\mathbb{R}}
  \newcommand{\C}{\mathbb{C}}

   \newcommand{\im}{\mathbf{i}}
    
   \newcommand{\abs}[1]{\left\vert#1\right\vert}
   \newcommand{\set}[1]{\left\{#1\right\}}

      \newcommand{\DOZZ}{\mathbf{C}}
   \newcommand{\tDOZZ}{\tilde{\mathbf{C}}}
   \newcommand{\Barnes}{\widetilde{G}}

\begin{document}
\title{{Log-correlated Random Energy Models with extensive free energy fluctuations:
	pathologies caused by rare events as signatures of phase transitions}}
\author{Xiangyu Cao}
\address{Department of Physics, University of California, Berkeley, Berkeley CA 94720, USA}
\address{LPTMS, CNRS (UMR 8626), Univ. Paris-Sud, Université Paris-Saclay, 91405 Orsay, France}
\author{Yan V. Fyodorov}
\address{King's College London, Department of Mathematics, London  WC2R 2LS, United Kingdom}
\author{Pierre Le Doussal}
\address{CNRS-Laboratoire de Physique Théorique de l'Ecole Normale Supérieure, 24 rue Lhomond, 75231 Paris, Cedex, France}
\begin{abstract}
We address systematically an apparent non-physical behavior of the free energy moment generating function for several instances of the logarithmically correlated models: the Fractional Brownian Motion with Hurst index $H = 0$ (fBm0) {(and its {\it bridge} version)}, a 1D model appearing in decaying Burgers turbulence with  log-correlated initial conditions, and finally, the two-dimensional logREM  introduced in [Cao et al., Phys.Rev.Lett.,118,090601] based on the 2D Gaussian free field (GFF) with background charges and directly related to the Liouville field theory.
All these models share anomalously large fluctuations of the associated free energy, with a variance proportional to the log of the system size. We argue that a seemingly non-physical vanishing of the moment generating function for some values of parameters is related to the termination point transition (a.k.a pre-freezing). We study the associated universal log corrections in the frozen phase, {both for log-REMs and for the standard REM}, filling a gap in the literature. For the above mentioned integrable instances of logREMs, we predict the non-trivial free energy cumulants describing non-Gaussian fluctuations on the top of the Gaussian with extensive variance. Some of the predictions are tested numerically.
\end{abstract}
\maketitle

\section{Introduction}
Fractional Brownian motions (fBm) were introduced first by Kolmogorov (in 1940) and later independently by Mandelbrot and van Ness. They are uniquely characterized as Gaussian random processes having zero mean, {stationary increments} and {self-similarity}. These properties determine a family of processes $B_{H}(x)$, parametrized by a \textit{Hurst exponent} $H$, describing the ``roughness'' of $B_H(x)$ or the  scaling of its increments:$
 \overline{(B_{H}(x) - B_{H}(x'))^2} \propto \abs{x-x'}^{2H} $. In particular, the Brownian motion (Wiener process) corresponds to $H = \frac{1}{2}$.  Yet, the limit $H \to 0_+$, which is one of the models we study in this work, does not make sense naively. A consistent way of defining a non-trivial extension of the fBm with $H=0_+$ (fBm0, denoted below as $B_0(x)$) was suggested in Ref.~\cite{fyodorov2016fbm}, and some statistics associated with the corresponding model was then investigated in Ref.~\cite{fyodorov2015moments}. In a nutshell, it was shown that fBm0 can be properly defined as a \textit{log-correlated process}, whose increments increase as the log of distance: $$\overline{(B_0(x) - B_0(x'))^2} =  \ln \frac{\abs{x-x}'^2 + \epsilon^2}{\epsilon^2} \,, $$
where a short-distance cutoff $\epsilon > 0$ is necessary to regularize the divergence of $\ln \abs{x-x'}$ when $x' \to x$. As it turns out, the singular short-distance behavior of fBm0, absent in $H>0$-fBm's, has important consequences, in particular for the extreme value statistics of the process.

To explain this point,  one may view the process $B_H(x)$ as a random energy potential, and consider the statistical mechanics model of a  particle thermalized in that potential. Such a model is defined by the associated partition function $ Z_B = \int \exp(-\beta B_H(x)) \dif x$ where $\beta$ is the inverse temperature. As discussed, e.g. in \cite{carpentier2001glass}, as long as $H > 0$ the associated Boltzmann-Gibbs probability weights $p_{\beta}(x)=Z_B^{-1}e^{-\beta B_H(x)}$ {\it at any temperature} $T=1/\beta>0$ are typically dominated by the absolute minimum of the potential $B_H(x)$ and its small neighborhood. In contrast, when $H = 0$, there is a non-trivial competition between the entropy and the deepest minima of $B_0(x)$; as a result, in such a system there is a \textit{freezing transition} at a finite critical inverse temperature $\beta=\beta_c$, which we can always ensure to be $\beta_c = 1$ by proper normalization (see \textit{e.g.}, Ref.~\cite{cao16order}, sect. 2.1).  Such freezing transition is not limited to fBm0, but is a general property of the class of log-correlated Random Energy Models (logREMs), \textit{i.e.}, the statistical mechanics models of a thermalized particle in a log-correlated random potential.  Models in this class arise in various contexts, \textit{e.g.}, spin glass theory~\cite{derrida1980random,derrida1988polymers,derrida16kppfinitesize,derrida2017finite}  extremal properties of branching processes~\cite{krapivsky00kpp,arguin2012poisson}, 2d XY model~\cite{carpenter98XY,carpentier00XYlong}, Anderson localization transitions~\cite{chamon1996localization,castillo97dirac,kogan96prelocalised}, random matrix and number theory~\cite{fyo12zeta,FyoKeat14,FyoSim15,ostrovsky2016riemann,arguin2017,arguin2017zeta}. In one and two spatial dimensions, log-correlated processes are akin to the 2d Gaussian Free Field (2d GFF), a fundamental object behind 2d conformal field theory, and therefore many logREMs are integrable by the replica approach and with the help of exactly solvable Selberg-type Coulomb gas integrals (in 1D)~\cite{fyodorov2008statistical,fyodorov2009statistical, fyodorov2010freezing,fyodorov2015moments,cao15gff,cao16maxmin,ostrovsky2016gff}, or by mapping to Liouville conformal field theory~\cite{cao16liouville,remy2017fyodorov,remy2017liouville,rhodes2017tail}.
Using these methods, it is sometimes possible to obtain exact predictions of observables such as the free energy distribution, Gibbs measure correlations, and in particular minimum value and position distributions.

Despite these apparent successes, in a few cases some {puzzling, even seemingly pathological/non-physical} features of the resulting expressions were noticed, mainly related to the free energy distribution in logREMs defined on unbound regions~\cite{fyodorov2009statistical,fyodorov2010freezing,cao16liouville}, as well as for  fBm0 restricted to the $[0,1]$ interval~\cite{fyodorov2015moments}. The main aim of the present paper is to suggest a way to re-interprete, and eventually cure these pathologies. A common feature of all the above cases is an anomalously large fluctuation of free energy: its variance is extensive in these models, while being of order unity in ordinary logREMs.  This observation turns out to be crucial for resolving the puzzles involved. Indeed, it will be clear that in the replica-trick approach, the resulting Coulomb gas integrals do not correspond to the free energy moment generating function as in ordinary logREMs, but to the \textit{non-Gaussian cumulant corrections} to a Gaussian distribution with extensive variance. We will argue that discarding such results as non-physical based on the observation that the non-Gaussian corrections cannot be the cumulants of a valid probability distribution is not at all warranted. Instead, after correct re-interpretation the corresponding expressions yield non-trivial predictions which can be tested numerically (and for some cases, are tested in this paper).

To illustrate our point, we first focus on the case of fBm0 in section~\ref{sec:fbm}.  For such a case the extensive free energy variance arises as a consequence of the fact that the random potential is pinned to $0$ at the origin \footnote{we are grateful for this observation pointed out to us by D. Ostrovsky.}. We show numerically that the non-Gaussian corrections to the free energy cumulants are correctly predicted by a standard replica-trick calculation. This answers \textit{positively} the question whether these formal results do have a statistical interpretation. Another known puzzle is related to a ``problematic zero'' of the analytically continued Coulomb gas integrals, observed in Ref.~\cite{fyodorov2015moments}. The latter paper pointed out rightly that it could be related to some phase transition. Here, we make this intuition more precise, by relating the problematic zero to {\it a termination point transition}~\cite{evers2008anderson,cao16liouville}, also known as the {\it pre-freezing}~\cite{fyodorov2009pre,wong17prefreezing}. Such a transition is known to be associated with additional log-correction factors and we extend the results of Refs.~\cite{cao16liouville,cao17thesis} for these corrections to any temperature in the Appendix~\ref{sec:free2} {not only  for log-REMs but also for the standard REM, filling a gap in the literature. }

Finally, in section~\ref{sec:ir} we apply the same approach to two logREMs defined on unbounded domains: the 1D Gaussian model~\cite{fyodorov2010freezing} which originally appeared in the problem of decaying Burgers turbulence with  log-correlated initial conditions, and finally the two-dimensional logREM introduced in Ref.~\cite{cao16liouville} which is based on the 2D Gaussian free field with background charges, and is directly related to the Liouville field theory and associated Dotsenko-Fateev Coulomb integrals. In particular, we predict the non-Gaussian cumulant corrections to the free energy distribution of that model, and discuss the problematic zero of the associated moment generating function, which we assign to yet another termination point transition (section~\ref{sec:DF}).

\section{fBm0 as pinned logREMs}\label{sec:fbm}
We first show that fBm0 can be defined as a  \textit{pinned logREM}. For this let $V_j, j = 1, \dots, M$ be an ``ordinary'' logREM discrete potential sequence with zero mean and logarithmically decaying correlations. We refer to Ref.~\cite{cao17thesis} (section 2.2.1) for a more precise definition. Here, we will concentrate on a few principle examples, which are all one-dimensional:
\begin{itemize}
\item[\textit{i.}] $V_j$ is the discrete potential of the circular model~\cite{fyodorov2008statistical}:
\begin{equation}
\overline{V_j} = 0 \,,\, \overline{V_j^2} = 2 \ln M  + w \,,\, \overline{V_j V_k} = - 2 \ln \abs{e^{2\pi\im j/M} - e^{2\pi\im k/M}} \,,\, \abs{k - j}\gg 1 \,.
\label{eq:circledef}
\end{equation}
Here and below $w$ denotes an $O(1)$ correction that depends on the model and $M$ (but has a calculable limit as $M\to \infty$). This logREM is obtained by restricting the (infinite-plane) 2d GFF to the unit circle, and is one of the most studied models in this class~\cite{fyodorov2008statistical,cao15gff,cao16maxmin,ostrovsky2016gff}. More precisely, we define the covariance matrix by discrete Fourier transform: $ \overline{V_j V_k} =  2 \sum_{p=1}^{M/2} \cos(2 \pi p (j-k) / M) / p$.
 \item[\textit{ii.}]  The interval model without charge~\cite{fyodorov2015moments,fyodorov2009statistical}:
 \begin{equation}
 \overline{V_j} = 0 \,,\, \overline{V_j^2} = 2 \ln M  + w \,,\, \overline{V_j V_k} =   2 \ln \frac{M}{\abs{k-j}}  \,,\,\abs{k - j}\gg 1  \,. \label{eq:intervaldef}
 \end{equation}
This logREM is obtained by restricting the same 2d GFF onto the interval $[0,1]$. {As a numerical remark~\cite{fyodorov2009statistical}, we note that although the continuum covariance matrix $C(x,y) = -2\ln \abs{x-y}$, $x,y\in[0,1]$ is not translationally invariant, fast Fourier transform can still facilitate its sampling. Indeed we can extend $C(x,y)$  to a cyclic covariance matrix for $x,y\in[0,2]$, \textit{viz}, $C(x,y) = -2\ln \left[ \min\left(\abs{x-y}, 2-\abs{x-y}\right) \right]$. Its Fourier expansion is $C(x,y) =  2 + 4 \sum_{p=1}^\infty \cos(\pi (x-y) p) \mathrm{Si}(\pi p) / (\pi p)$,  where $\mathrm{Si}(x) = \int_0^x \sin(y) y^{-1} \dif y$ is the sine integral. A discrete version can be obtained by replacing $x,y=j/M, k/M$ and cutting off the sum up to $p=M$.}
%\red{if you don't know this, it's new. So I explain it a bit.}
 \end{itemize}

Now, given any 1d logREM and a marked point, which we fix as $j=1$, we define the corresponding  \textit{pinned logREM} by the potential:
\begin{equation}
B_j := V_j - V_1  \,. \label{eq:pinneddef}
\end{equation}
This pins the value of the potential sequence $B_j$ at $j = 1$ to $B_1 = 0$. The covariance matrix of $B_j$ is simply related to that of $V_j$ as follows:
 we have $\overline{B_j B_k}^c = C_{jk} - C_{j1} - C_{1k} + C_{11}$, where we denoted $C_{jk} := \overline{V_j V_k}^c$.  Note that $B_j$ has also zero mean: $\overline{B_j} = 0$ for any $j$.

Let's consider the example where $V_j$ is the potential of the interval model.  In that case $C_{jk} = C(\abs{j-k})$ depends only on the distance, so that $C(0) = 2 \ln M$ and $C(j) = 2 \ln M - 2 \ln(j)$ for $M > j \gg 1$. By the relation~\eqref{eq:pinneddef} the increments' covariance structure is then given by
\begin{equation}
\overline{(B_j - B_k)^2}^c = \overline{(V_j - V_k)^2} = 2 C(0) - 2 C(\abs{j-k}) \stackrel{\abs{j-k} \gg1}= 4 \ln \abs{j-k}
\end{equation}
We see that the increments are stationary, and the variance grows logarithmically. Combined with $B_1 = 0$, this pinned interval model qualifies as a definition of a {discrete version} of the fBm0, $B_{H=0}$.

In the example where $V_j$ is the potential of the circular model, the increments of $B_j$ are also stationary: $\overline{(B_j - B_k)^2}^c = 2C(\abs{j-k}) \stackrel{\abs{j-k} \gg1}= 4 \ln \sin ( \pi \abs{j-k}/ M)$, and $B_1 = 0$. So the pinned model defines a periodic fBm0 that starts at and returns to $0$, {hence can be called a {\it fBm0 bridge}}.

\subsection{Free energy: Large deviation function and termination point transition}\label{sec:free1}
\begin{figure}
\includegraphics[scale=0.6]{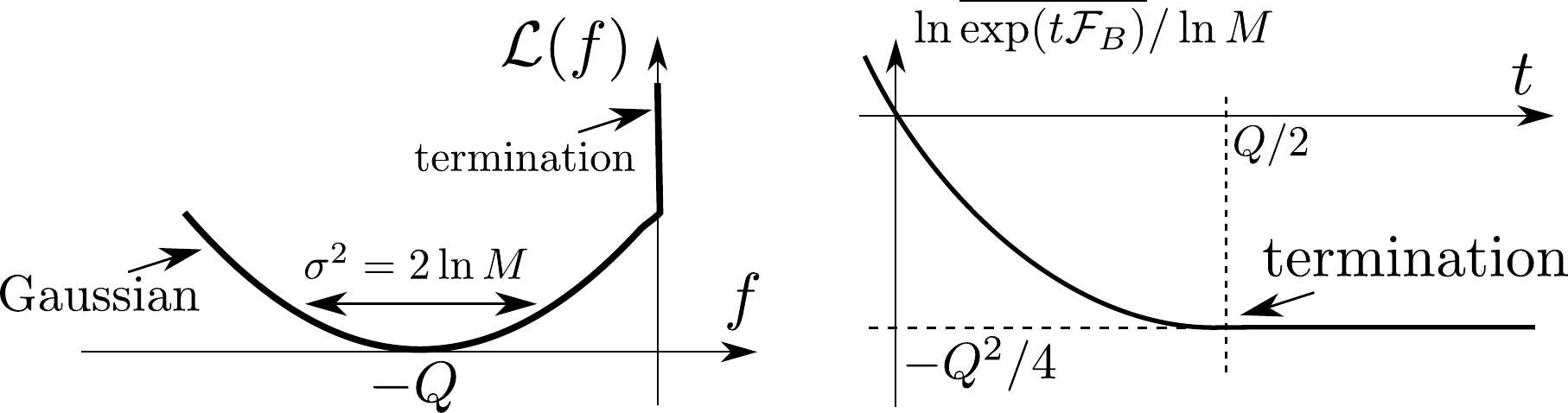}
\caption{ \textit{Left}: large deviation rate function of the free energy $ \mathcal{F}_B $, eq. \eqref{eq:ldf}. The variance $\sigma^2$ refers to that of $\mathcal{F}_B$ \textit{Right}:  the leading exponent of the moment generating function of $ \mathcal{F}_B$, eq. \eqref{eq:termination}. The two functions are related by Legendre transform.}\label{fig:Lf}
\end{figure}
{Although many quantities of interest related to the Gibbs measure for the above pinned logREMs can be successfully evaluated, it remains}  an open challenge~\cite{fyodorov2015moments} to calculate the {associated} free energy distribution, defined by the partition function:
\begin{equation} \mathcal{Z}_B = \sum_{j=1}^M e^{-\beta B_j} \,.  \label{eq:ZB} \end{equation}
By the definition eq. \eqref{eq:pinneddef}, we can relate $\mathcal{Z}_B$ directly to the Gibbs weight $p_{\beta,j}$ of the logREM with potential $V_j$ (which we shall call the ordinary logREM):
\begin{equation}\label{eq:ZBisp}
\mathcal{Z}_B = \mathcal{Z} e^{\beta V_1} = p_{\beta,1}^{-1}  \quad \text{ where } \quad p_{\beta,j} :=  \mathcal{Z}^{-1}e^{-\beta V_j} \,,\,
\mathcal{Z} = \sum_{j=1}^{M} e^{-\beta V_j } \,.
\end{equation}
{In this way we have reduced the problem of finding the free energy distribution of a pinned logREM/fBm0 model to that of the Gibbs weight $p_{\beta,1}$ of the ordinary logREM:}
\begin{equation}\mathcal{F}_B = -\beta^{-1}\ln\mathcal{Z}_B = \beta^{-1} \ln p_{\beta,1} \label{eq:FBmulti}\end{equation}

{As is well-known, the Gibbs measure associated with ordinary logREMs is {\it  multifractal} \cite{chamon1996localization,castillo97dirac,kogan96prelocalised,fyodorov2009pre} which is reflected in the nontrivial scaling
of moments of the Gibbs weights $p_j$ with the system size $M$. }
 At the leading order, the associated large deviation function of $\mathcal{F}_B$ is therefore directly related to the multifractal spectrum of the Gibbs measure. As a result, we have, {for large $M\gg 1$}:
\begin{equation}
\mathcal{L}(f)  := -\ln \text{Prob.}(f = \mathcal{F}_B / \ln M) / \ln M = \begin{dcases}
(f + Q)^2 / 4 & f < 0 \\
+ \infty & f \geq 0
\end{dcases} \,,\, Q = b + b^{-1} \,,\, b = \min(1, \beta) \,. \label{eq:ldf}
\end{equation}
 {See Fig.~\ref{fig:Lf} for an illustration.}

A simple argument to understand the above result is the following~(\cite{cao17thesis}, sect. 2.1.4). Eq. \eqref{eq:ZBisp} implies that
$ \mathcal{F}_B = \mathcal{F} - V_1$, where $\mathcal{F} = -\beta^{-1}\ln \mathcal{Z} $ is the free energy of the ordinary logREM. Its universal extensive behavior was predicted in ~\cite{carpentier2001glass}:
\begin{equation}
\mathcal{F}  = -Q \ln M + \chi \ln \ln M + O(1)\,,\, \label{eq:freezing}
\end{equation}
where $\chi = \frac32$ when $\beta > 1$ ({see Ref.~\cite{rosso12counting,fyodorov2015high} for a universal mechanism behind this exponent in disordered multi-fractals, and  Ref.~\cite{ding2014extreme} for a rigorous proof for a general log-correlated fields}), $\frac12$ when $\beta = 1$ and $0$ when $\beta < 1$. {The  log-corrections are universal and are closely associated with the logREM freezing transition, whereas an $O(1)$ is the non-universal fluctuating part of the free energy, whose variance is of order unity. Thus  $\mathcal{F} / \ln M = -Q + o(\ln M)$ in a typical realization.  On the other hand, $V_1$ is a Gaussian variable of zero mean and variance $2\ln M$. Therefore, one should expect that $f:= \mathcal{F}_B / \ln M$ is a Gaussian variable with mean $-Q$ and variance $2/\ln M$.} This leads to the large deviation function~\eqref{eq:ldf}  for $f < 0$.

However, the same expression cannot be valid when $f > 0$. {Indeed, since the Gibbs  weight $p_{\beta,1} \leq 1$, $\mathcal{F}_B = \beta^{-1} \ln p_{\beta,1}$ can never be positive. This fact is precisely behind the ``hard wall'' condition in the bottom line of eq.~\eqref{eq:ldf}, the value at $f = 0$ being the so-called \textit{termination point} of the Gibbs measure multi-fractal spectrum. The realizations where $f \sim 0$ are rare (since typically $f \sim -Q$), and are such that the ``pinned'' value $B_1 = 0$ is amongst the deepest minima of the potential $B_j$. This implies that the Gibbs probability weight $p_{\beta,1}$ in these realizations is of order unity. Note that as the points near $1$ contribute significantly to the free energy, {\color{green} the values of}  $V_1$ and $\mathcal{F}$ (and thus $\mathcal{F}_B$) become strongly correlated when conditioned to these realizations.}

Let us stress the most important difference between fBm0 models and ordinary logREMs:  $\mathcal{F}_B$ has an \textit{extensive} variance $2\ln M + O(1)$, whereas for usual logREMs the variance is of { order} $O(1)$. { As such a feature seems to be at the heart of the peculiarities of the models that we are considering, we give a simpler and rigorous verification of this fact for the particular case of the circular model. Since without pinning  the potential is statistically translationally invariant, we have $\overline{\mathcal{F} V_1}^c = M^{-1} \overline{\mathcal{F} \sum_{j=1}^M V_j}^c = 0$,  since the ``zero-mode'' $ \sum_{j=1}^M V_j$ vanishes in all realizations [see below eq.~\eqref{eq:circledef}].} Therefore, as $\mathcal{F}_B = \mathcal{F} + V_1$, we further have
\begin{equation}\label{eq:variance1}
\overline{\mathcal{F}_B^2}^c = \overline{V_1^2} + \overline{\mathcal{F}^2}^c =  2 \ln M + { w} + \overline{\mathcal{F}^2}^c  \,,
\end{equation}
where the two last terms are of order unity, and known exactly~\cite{fyodorov2008statistical} (see also eq.~\eqref{eq:cumulants} below).

Finally, let us discuss the moments of the partition function (or the moment generating function of the free energy) $\overline{\mathcal{Z}^n} = \overline{\exp(t \mathcal{F}_B)}$, $t = -n\beta$. Those are directly given by the moments of the Gibbs weight \textit{via} eq.~\eqref{eq:ZBisp}, and thus closely related to the so-called inverse participation ratios in the ``annealed'' ensemble, see  further detail and discussion in Ref.~\cite{fyodorov2009pre}. {The leading large-$M$ behavior of such a generating function} is then obtained as the Legendre transform of the large deviation function given in eq. \eqref{eq:ldf}. The result is~\cite{castillo97dirac} {(see Fig.~\ref{fig:Lf} for an illustration)}:
\begin{equation}
 \overline{e^{t\mathcal{F}_B}}  = \overline{p_{\beta,1}^{t/\beta}} =  \begin{dcases}
 M^{-t Q + t^2 + o(1)} &  t < Q/2 \\
 M^{-Q^2/4 + o(1)} & t \geq Q/2
 \end{dcases} \,, \label{eq:etfleading}
\end{equation}
where $o(1)$ denotes finite-size corrections that go to zero in the $M \to \infty$ limit. The hard wall, or termination point, at $f = 0$  gives rise to a non-analytic behavior of the leading scaling exponent at $t = Q/2$, known as the {\it termination point transition}, also known as {\it pre-freezing}~\cite{fyodorov2009pre,wong17prefreezing}. {Beyond that point the exponent ``freezes'', i.e. becomes independent of $t$, similarly to the free energy density $\mathcal{F}/\ln M$, which also ``freezes''   beyond $\beta = 1$}.  The similarity between termination point and freezing transition goes beyond the leading order: the {multiplicative} log-corrections to eq.~\eqref{eq:etfleading}  turn out to be reminiscent to those of the freezing transition. Such corrections were predicted in Refs.~\cite{cao16liouville,cao17thesis} in the high-temperature $\beta < 1$ phase. In the Appendix \ref{sec:free2} we extend these results to any temperature.

\subsection{Coulomb gas integrals}\label{sec:free3}
The predictions in the previous section (and in the Appendix~\ref{sec:free2}) are expected to be universally valid for all logREMs in the thermodynamic limit $M \to \infty$. For a few integrable logREMs, we may go further to predict the precise value of $O(1)$-terms above. We shall first focus on the example of the circular model/periodic fBm0, defined in eq. \eqref{eq:circledef}; analytical results on the interval model were obtained in Ref.~\cite{fyodorov2015moments,ostrovsky2016gff} by relying upon the Selberg Coulomb gas integrals \cite{forrester2008importance,forrester2010log} and will be recalled briefly below.

\subsubsection{{Circular model and fBm0 bridge}}\label{sec:cir}
The approach of this section is  based on employing the standard heuristic method of the physics of disordered systems known as the
replica trick. Roughly speaking, it starts by considering partition function integer moments $\overline{\mathcal{Z}_B^n}$, which, when $n = 1, 2, 3, \dots$, can be expanded as a sum over $n$ replica positions:
\begin{equation} \overline{\mathcal{Z}_B^n} = \sum_{j_1=1}^M \dots   \sum_{j_n=1}^M \overline{\exp\left( n \beta V_1 - \beta V_{j_1} - \dots -  \beta V_{j_n} \right)} \,,\, \label{eq:replica0} \end{equation}
where we used eq. \eqref{eq:ZB} and eq. \eqref{eq:pinneddef}. Note that the disorder average can be simply performed by Wick theorem (using eq. \eqref{eq:circledef} for the circular model). Then one replaces the sum by a Coulomb gas integral in the thermodynamic limit; when the integral has an exact expression, one can analytically continue it to arbitrary complex $n$~\cite{fyodorov2008statistical,fyodorov2010freezing,ostrovsky2009mellin} and obtain $\overline{\exp(t\mathcal{F}_B)}$ for generic $t$.
{The correspondence between discrete sums and continuum integrals is determined by the {replica symmetry breaking mechanism (RSB) which may or may not be operative in the phase in question, and so depends} on whether $\beta < 1$ and $t < Q/2$. For ordinary logREMs, which in their free energy only exhibits a freezing transition, the formalism is described in Refs.~\cite{fyodorov2010freezing,cao16order,cao17thesis}.  In the present case of pinned logREM, the termination point/prefreezing transition is also present, and requires extending the RSB formalism; such an analysis was initiated in Ref.~\cite{fyodorov2009pre} and developed in further generality more recently in Ref.~\cite{cao17seiberg} (in particular, section 2 and Appendix B), from which we shall apply some results.

 Let us start within the (high temperature) phase where $\beta < 1$ and $t < Q/2$, so that the replica symmetry is unbroken. The sum eq. \eqref{eq:replica0} can be replaced by integral in a naive way, \textit{i.e..},  the moment generating function of $\mathcal{F}_B$ is given  in the $M \to \infty$ limit by a Coulomb gas integral:
\begin{equation}
\overline{\exp(t \mathcal{F}_B)} = M^{-Qt + t^2} e^{\frac12 w (t^2 - t)} \mathbf{M}(n = -t / \beta, a = t, b = \beta) \,, \label{eq:continuous}
\end{equation}
where $\mathbf{M}$ is known as the Morris integral, defined as~\cite{forrester2008importance}:
%{Equation has been corrected please recheck}
\begin{align}
\mathbf{M}(n, a, b) &= \int_{0}^{2\pi} \prod_{i=1}^{{n}} \left[ \frac{\dif\theta_i}{2\pi}  \abs{1 - e^{\im \theta_i}}^{-2 a b} \right]
\prod_{i<j} \abs{e^{\im \theta_i} - e^{\im \theta_i}}^{-2b^2}  = \prod_{j=0}^{{n}-1} \frac{\Gamma(1 - 2 a b-jb^2)\Gamma(1-(j+1)b^2)}{\Gamma(1 -  a b - jb^2)^2 \Gamma(1 - b^2)}     \\
& {= \frac{ \tilde{\mathbf{M}}(n, a, b)}{\Gamma^{{n}}(1-b^2)} \quad \text{where} \quad
 \tilde{\mathbf{M}}(n, a, b)
=\Gamma(1 - {n}b^2)} \frac{\Barnes_b(Q-2a)\Barnes_b(Q - a -{n}b)^2}{\Barnes_b(Q-2a - nb)\Barnes_b(Q - a)^2} \frac{{\Barnes_b(Q)}}{\Barnes_b(Q - n b)} \label{eq:morrisgamma}
\end{align}
{Here $\Barnes_b$ is the generalized Barnes function. Its defining property is the following functional relation [we adopt the notation of Ref.~\cite{fyodorov2015moments},  see eq. (237) therein]
\begin{equation}  \Gamma(b x ) = \frac{ \Barnes_b(x + b) }{\Barnes_b(x)} \quad \Leftrightarrow   \quad \prod_{j=1}^{n} \Gamma(b x -j b^2) = \frac{ \Barnes_b(x) }{\Barnes_b(x - n b)} \,, \label{eq:functional} \end{equation}
which facilitates the analytically continuation of products of Gamma functions. $\Barnes_b(x)$ is an entire function with the following simples zeros:
\begin{equation}
\Barnes_b(x) = 0 \,,\,  x = -n b - m/b \,,\, n,m = 0,1,2,\dots\,. \label{eq:Barneszeros}
\end{equation}
  When $b = 1$, $\Barnes_b(x)$ reduces to the ordinary Barnes function:
\begin{equation}
\Barnes_1(x)=G(x) = { (2 \pi)^{x/2} \exp\left( (x-1)(\log\Gamma(x) - x/2) - \psi^{(-2)}(x)\right)} \label{eq:Barnes}
\end{equation}
where $\psi ^{(n)}(x)$ is $n$-th poly-gamma function. Note that the first line of eq.~\eqref{eq:morrisgamma} holds only when the integral converges, whereas the second line is an analytical continuation that makes sense for general complex value of parameters. We refer to D. Ostrovsky's work on rigorous aspects of such a procedure~\cite{ostrovsky2009mellin,ostrovsky2013selberg,ostrovsky2013theory,ostrovsky2016barnes,ostrovsky2016gff}.

Now, in the phase defined by  $\beta > 1$, $t < Q/2=1$ [note that in the $\beta > 1$ phase, $Q = 2$, see eq.~\eqref{eq:ldf}], the above expression is modified by the freezing transition in a fashion known as the duality-freezing scenario (which can be understood by a breaking of replica symmetry~\cite{fyodorov2010freezing,cao16order} {occurring in the bulk and unrelated to the presence of the pinning at a particular point}), and becomes (with $Q=2$)
\begin{equation}
\overline{\exp(t \mathcal{F}_B)} \Gamma(1 + t/\beta) = M^{-2t + t^2 + c t} e^{\frac12 w t^2}  \Gamma(1 + t)
\tilde{\mathbf{M}}(n = -t, a = t, b = 1) \,, \label{eq:etFBMorris}
\end{equation}
where $c = \frac32 \ln \ln M / \ln M + c_{\text{UV}}$ contains the log-correction of eq. \eqref{eq:freezing} and the constant $c_{\text{UV}}$ that depends on the short-distance details of the model~\cite{cao16order}. In particular, by expanding the above equation at $t = 0$, we obtain the cumulants of $\mathcal{F}_B$. At zero temperature, we thus obtain that the cumulants of {the distribution of the minimum $B_{\min}$ for the fbM0 bridge, $B_j$, are a sum of those of a Gaussian distribution of variance $2 \ln M$ and \textit{non-Gaussian corrections}, whose values are given in the $M\to\infty$ limit as (with $Q=2$)
\begin{align}
&\overline{B_{\min}^2}^c - 2 \ln M - w \stackrel{M\to\infty}{\longrightarrow} C_2  \,,\,  \overline{B_{\min}^k}^c \stackrel{M\to\infty}{\longrightarrow}  C_k \,,\, k > 2 \,;\,  \\
&
C_k := \left. \frac{\dif^k}{\dif t^k} \ln \left[\Gamma(1 + t) \tilde{\mathbf{M}}( -t,  t, 1)  \right] \right\vert_{t=0} \,
=  \left. \frac{\dif^k}{\dif t^k} \ln \left[ \frac{G(2-2 t) \Gamma(1 + t)^2 }{G(2-t)^3 G(2+t)}
 \right] \right\vert_{t=0} \label{eq:cumulants0} \\
&\set{C_2, C_3, C_4}_{\text{circular}} = \set{\frac{\pi ^2}{3},-2 \pi ^2+ 8 \zeta(3),\frac{14 \pi ^4}{15}-72 \zeta (3)} = \set{3.28987, -10.1228, 4.36705} \,. \label{eq:cumulants}
\end{align}}
Here  $\zeta(x)$ is the Riemann zeta. The above predictions are tested numerically, see Fig.~\ref{fig:fbm} (a)
{The prediction for $C_2$ is tested by computing $\overline{B_{\min}^2} - \overline{V_1^2}^c$.}
As an independent check, we recall that $C_2 = \frac{\pi ^2}{3}$ is known as the minimum variance  of the circular model without pinning~\cite{fyodorov2008statistical,fyodorov2009statistical}. Thus, we recover eq. \eqref{eq:variance1}, which was obtained rigorously.
{Higher cumulants $C_k$ are also easily expressed in terms of poly-gamma functions,
using the formula Eq. \eqref{phik} in the Appendix.}

\begin{figure}
	\includegraphics[scale=.45]{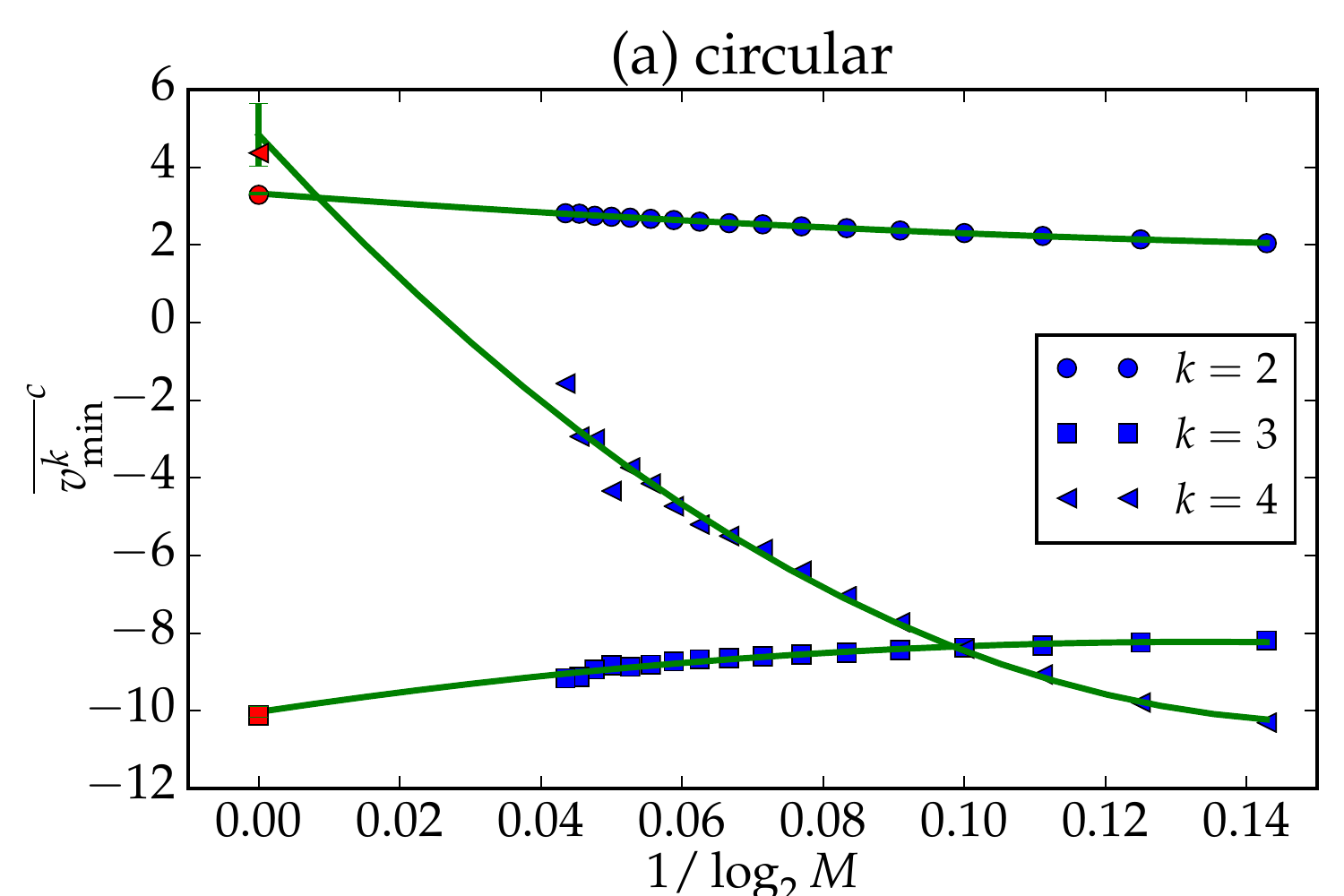}
		\includegraphics[scale=.45]{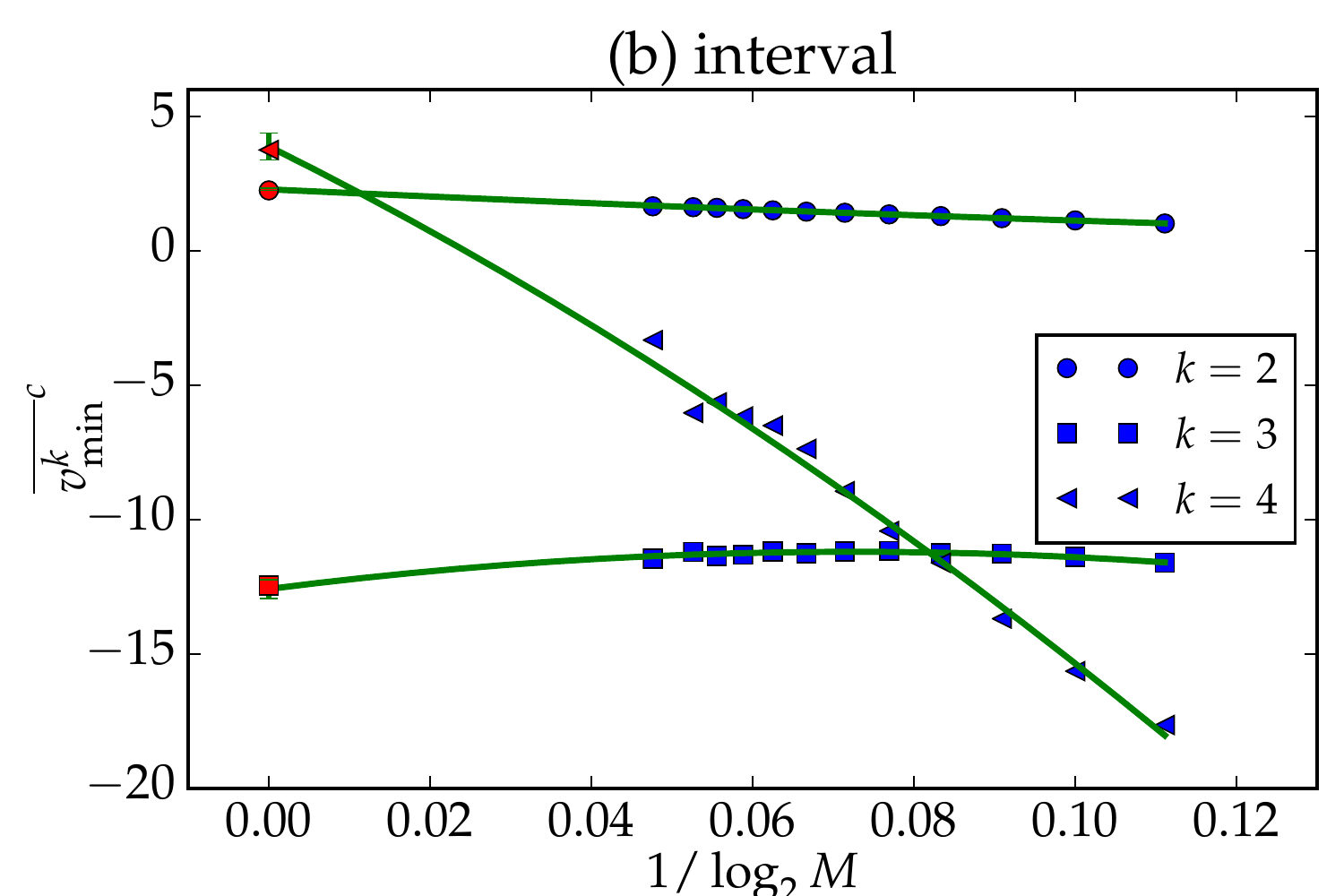}
	\caption{(a) Numerical calculation of the cumulants of the minimum {of the fBm0 bridge.} For the variance, $k = 2$, we define {$\overline{v_{\min}^2} :=  \overline{B_{\min}^2} - \overline{V_1^2}^c$}, {\textit{i.e.}, the extensive Gaussian contribution is subtracted from the raw data. Higher cumulants are not affected by the Gaussian contribution, and are plotted as such: {$\overline{v_{\min}^k} :=  \overline{B_{\min}^k}$}, $k>2$.} A quadratic Ansatz $a_1 + a_1 / \ln M + a_2 / (\ln M)^2$ is used to account for the finite size scaling and extrapolate the $M \to \infty$ value from $M = 2^{8} \to, 2^{23}$, which is compared to the predictions eq. \eqref{eq:cumulants} (in red markers). (b) The same for the fBm0 on the interval $[0,1]$ (pinned interval model, $M = 2^{10}, \dots, 2^{23}$), compared with eq. \eqref{eq:cumulants2}. } \label{fig:fbm}
\end{figure}
In general, at any temperature, the free energy $\mathcal{F}_B$'s cumulants are the sum of those of a Gaussian of variance $2\ln M$ and non-Gaussian corrections $C_{k,\beta}$, which are given by the Taylor expansion of the analytically continued Morris integral {with $Q=2$}
%\red{same remark as above, see also the tilde on M}
\begin{align}
& \overline{\mathcal{F}_B^2}^c = 2 \ln M + w + C_{2,\beta} \,,\,  \overline{\mathcal{F}_B^k}^c = C_{k,\beta} \,,\, k > 2\,;   \\
&C_{k,\beta}=\left.  \frac{\dif^k}{\dif t^k} \ln \left[\tilde{\mathbf{M}}( -t / \beta, t, \beta) \right]\right\vert_{t=0} \,,\, \beta < 1 \,,\, k > 1 \,;  \\
& C_{k,\beta} = C_{k} - \beta^{-k} (-1)^k (k-1)! \zeta(k)  \,,\, \beta > 1 \,, k > 1\,. \label{eq:Cbetaklow}
\end{align}
We emphasize that the last identity is a direct consequence of the freezing scenario, and applies to the low-temperature phase of \textit{all} models  in this work, as well as ordinary logREMs [in which case it was known since Ref.~\cite{fyodorov2009statistical}, eq. (24)].
Notice that the low-temperature variance $C_{2,\beta} = C_2 - \pi^2/(6\beta^2)$ is smaller than the zero-temperature one. Heuristically speaking, this reflects the fact that the non-Gaussian fluctuations of the free energy of logREMs in the frozen phase are dominated by those of the minimal energy.
For the fBm0 bridge case, eq.~\eqref{eq:Cbetaklow} and eq.~\eqref{eq:cumulants} imply more explicitly:
	\begin{eqnarray}
	C_{2,\beta} = \frac{(2 \beta^2-1) \pi^2}{6 \beta^2} \quad , \quad
	C_{3,\beta} = 2 (4 + \frac{1}{\beta^3}) \zeta(3) - 2 \pi^2 \,,\, \beta > 1 \,.
	\end{eqnarray}
Analogous formulas for the other models in the sequel can be similarly obtained and will not be displayed explicitly.

\subsubsection{fBm0 on an interval}
The method above applies also to the interval model (or fBm on $[0, 1]$), defined in eq. \eqref{eq:intervaldef}, upon replacing the Morris integral $\mathbf{M}(n, a, b)$ by a special case of the Selberg integral~\cite{fyodorov2015moments}:
\begin{align}
\mathbf{S}(n,a,b) &:= \int_0^1 \prod_{i=1}^{n} \left[x_i^{-2ab} \dif x_i  \right] \prod_{i<j} \abs{x_i - x_j}^{-2b^2}  =
\prod_{j=0}^{n-1} \frac{\Gamma(1-2ab-jb^2) \Gamma(1 - j b^2) \Gamma(1-(j+1)b^2)}{\Gamma(2-2ab-(n+j-1)b^2)\Gamma(1-b^2)}  \nonumber \\
& {= \frac{ \tilde{\mathbf{S}}(n, a, b)}{\Gamma^{{n}}(1-b^2)} \quad \text{where} \quad
 \tilde{\mathbf{S}}(n, a, b)
= \frac{\Barnes_b(1/b) \Barnes_b(Q - 2 a) \Barnes_b(Q) \Barnes_b(2 Q - 2a-2nb)}{\Barnes_b(1/b - n b) \Barnes_b (Q-2a - n b) \Barnes_b(Q-n b)  \Barnes_b(2 Q - 2a-nb)} }   \,,\label{eq:selberg}
\end{align}
{which coincides with Eq. (238) in \cite{fyodorov2015moments} setting $\bar a=-2 a =2 b n$ and $\bar b=0$ there.}
We remark that Morris integral and Selberg integral (in their respective general form) are  related~\cite{forrester2008importance} and this fact has been used in Refs~\cite{fyodorov2015moments,ostrovsky2016gff}.  {We then obtain, following similar steps as above  the non-Gaussian corrections to the cumulants of the probability distribution for the mininum $B_{\min}$ of the $[0,1]$-fBm0, for $k>1$ [compare with eq.~\eqref{eq:cumulants} above]
\begin{eqnarray}
&& \overline{B_{\min}^2}^c - 2 \ln M - w \stackrel{M\to\infty}{\longrightarrow} C_2  \,,\,  \overline{B_{\min}^k}^c \stackrel{M\to\infty}{\longrightarrow}  C_k  \,, \\
&&
C_{k} = \left. \frac{\dif^k}{\dif t^k} \ln \left[ \Gamma(1 + t)
\tilde{\mathbf{S}}( -t, t, 1) \right]\right\vert_{t=0}
= \left. \frac{\dif^k}{\dif t^k} \ln \left[
\frac{2 G(2-2t) \Gamma(1+t) }{G(2-t) G(4-t) G(1+t) G(2+t)}
 \right]\right\vert_{t=0} \label{BB}  \,,
\end{eqnarray}
see also Eq. (236) in \cite{fyodorov2015moments}. Now we can go further than Ref.  \cite{fyodorov2015moments} and obtain the following explicit prediction for the lowest cumulants}
\begin{equation}
\set{C_2, C_3, C_4}_{[0,1]} = \set{\frac{9}{4},8 \zeta (3)-\frac{8 \pi ^2}{3}+\frac{17}{4},-72 \zeta (3)+\frac{4 \pi ^4}{5}+\frac{99}{8}} = \set{2.25, -12.4525, 3.75418} \,. \label{eq:cumulants2}
\end{equation}
These predictions are also found to be in nice agreements with numerical calculation, with similar strong finite size corrections, see Fig.~\ref{fig:fbm} (b). Note that since the interval model is not translationally invariant, the argument leading to eq.~\eqref{eq:variance1} is invalid and $C_2$ is considerably different from the variance of the minimum of the unpinned interval model, $ \frac{4 \pi ^2}{3}-\frac{27}{4} = 6.40947$~\cite{fyodorov2009statistical}. {Higher cumulants $C_k$ are also easily expressed in terms of poly-gamma functions,
using the formula Eq. \eqref{phik} in the Appendix.}

We now address two pathological features noticed in Ref.~\cite{fyodorov2015moments}, p57, and argue that they do not invalidate the foregoing predictions.

First, it was observed that the cumulant corrections $C_k$ cannot correspond to a well-defined positive probability distribution. For the sake of argument, let us call $v_{\min}$ a \textit{fictitious} random variable whose $k$-th cumulant is $C_k$. Then its fourth moment obtained from eq. \eqref{eq:cumulants2} would be negative: $\overline{v_{\min}^4} = -0.115 \dots< 0$. This feature would become problematic if and only if one wanted to view the minimum {$B_{\min}$ as a sum of a Gaussian of variance $2 \ln M$
and an independent random variable $v_{\min}$. We stress that such a ``natural'' interpretation of the above results eq.~\eqref{eq:cumulants}, \eqref{eq:cumulants2} is not possible in any way. However this does not
preclude the fact that the non-gaussian random variable
%can be only viewed as a formal short-hand: the ill-defined-ness of the fictitious $v_{\min}$ does not affect the fact that
$B_{\min}$ has a probability distribution whose cumulants are correctly predicted by the above equations~\eqref{eq:cumulants} and \eqref{eq:cumulants2}.

A second disturbing feature is that in both models studied so far, the moment generating function of $\mathcal{F}_B$ has a zero at $t = Q/2$, coming from the factor $\Barnes_b(Q - 2a)$ (with $a=t$) present in eq. \eqref{eq:selberg} and \eqref{eq:morrisgamma}, respectively [see eq.~\eqref{eq:Barneszeros}]. So the cumulant generating function $\ln \overline{\exp(t\mathcal{F}_B)}$ must become non-convex when $t$ is close enough to $Q/2$ [despite the presence of the $M^{t^2}$ factor in eq.~\eqref{eq:etFBMorris}], which calls for a further explanation. % As a consequence, it cannot be convex in the whole interval $t \in (0, Q/2)$. One would argue that since $\overline{\exp(t\mathcal{F}_B)} =  M^{t^2 - t c_\beta} m(t)$ where $m(t)$ is some $M$-independent fu for $M\gg 1$, for any \textit{fixed} $t \in (0, Q/2)$, when $M$ is big enough, $\ln \overline{\exp(t\mathcal{F}_B)}$ becomes convex at $t$. Nevertheless, for any fixed $M$, the convexity will fail at $t$ becomes close enough to $Q/2$, which calls for a further explanation

Nicely, the required explanation of this feature is provided by the considerations of section~\ref{sec:free1}: $t = Q/2$ is the locus of the termination point transition. Beyond that point and in the phase dominated by the termination/pre-freezing mechanism, the RSB becomes non-trivial: a finite portion of replicas become bound and freeze at $j \sim 1$~\cite{fyodorov2009pre,cao17seiberg}.
%\red{add reference to Yan prefreezing paper} \uv{added}.
For this reason, the {moment generating function of the discrete model} is not naively given by the continuum integral as indicated by eq.~\eqref{eq:continuous}. As argued in Ref.~\cite{cao17seiberg}, the zero at $t = Q/2$ is intimately related to the log-corrections associated with the termination point transition. Nevertheless, these modifications do not affect the validity of the free energy cumulant predictions eqs.~\eqref{eq:cumulants} and \eqref{eq:cumulants2}, which are determined by the derivatives at $t=0$ of the cumulant generating function and therefore is unaffected by a far zero at $t = Q/2$. We conclude that when the termination point
dominates in the large deviation regime of the free energy, it cannot affect the cumulant of the free energy
at leading order in the $M\to\infty$ limit.

\section{1D Gaussian and 2D logREM models}\label{sec:ir}
As we mentioned in the introduction, a few other logREMs defined on {\it unbounded} domains exhibit similar pathologies in their free energy cumulant generating function. As we shall see the reasons for this behavior are similar to the ones discussed above for the fBm0 models (pinned logREMs).
Enlightened by the understanding of these previous examples, let us review two more representative cases: the 1d Gaussian model~\cite{fyodorov2009statistical,fyodorov2010freezing,cao17thesis}, and the 2d GFF model in the plane in presence of two background charges. The latter was studied in Ref.~\cite{cao16liouville}, which showed that moments of the associated Gibbs probability density at any point in space can be mapped to four-point correlation functions of the Liouville field theory. Here we shall call this model simply {\it the 2D logREM}, since it is presently the only one in this class for which exact results are available. Its study is facilitated by using the famous Dotsenko-Fateev (DF) Coulomb gas integrals (see below). Note that the ``pathological features'' of these models concern uniquely their free energy fluctuation, not their Gibbs measure. The latter has been well understood in the above quoted works, so we shall focus on the former. For the sake of simplicity, we switch to the continuum formalism, and work in the  ``simple scaling'' part of the high temperature phase, i.e. $\beta < 1$, $t < Q/2$, unless otherwise stated.

\subsection{Gaussian model}\label{sec:gaussian}
The Gaussian model describes a disordered potential $V(x)$ on a 1d infinite line that is the sum of a parabola $\frac{x^2}{2}$ and the restriction of the 2d GFF on the real line $\phi(x)$:{
\begin{equation}
V(x) = \phi(x) + \frac{x^2}{2} \,.
\end{equation}}
 Its arises in the study of decaying Burgers equation in 1d with log-correlated initial data~\cite{fyodorov2010freezing}. To obtain a well-defined statistical model, one needs a large-distance cut off $L$ (in additional to a short distance cut off $\epsilon$ needed for all logREMs); the continuum partition function has the following form
\begin{equation}
\mathcal{Z}_G = \int_{-L/2}^{L/2} \frac{\dif x}{\sqrt{2 \pi} \epsilon} \exp(-\beta \phi(x) - \beta x^2/2) \,, \overline{\phi(x)} = 0 \,,\, \overline{\phi(x)^2} = 2 \ln (L/\epsilon) \,,\, \overline{\phi(x) \phi(x')} = 2\ln\abs{\frac{L}{x-x'}} \,,\,   \abs{x - x'} \gg \epsilon \,.
\end{equation}
When $L \to \infty$, %the replicated averages, i.e.
the positive integer moments of $\mathcal{Z}_G$, are exactly computable, thanks to the Mehta integral~\cite{forrester2008importance}. After analytical continuation using the Barnes function, we have, in the $\beta < 1$ phase (and in the $\epsilon\to0, L\to \infty$ limit):
%\red{why is there a $\sqrt{\beta}$ for each $L$ factor?} \uv{should not be there! corrected here and below}
\begin{equation}\label{eq:mehta}
\overline{\exp(t \mathcal{F}_G)} = \exp\left[t C_1 + t^2 \ln \left(L \right) \right] \frac{\Barnes_\beta(1/\beta)}{\Barnes_\beta(t + 1/\beta)} \,,\,
\end{equation}
where $C_1 = Q \ln \left(  \epsilon \sqrt{\beta} \right) + \ln \Gamma(1-\beta^2)^{\frac1\beta}$ is unimportant for the following. Implementing the freezing scenario (or RSB) in the $\beta > 1$ phase, we obtain in particular the following cumulant predictions for the minimum of the total potential $V(x)$ in the thermodynamic ($L\to\infty, \epsilon\to0$) limit:
\begin{equation}\label{eq:vminmehta}
\overline{V_{\min}^2}^c =  2 \ln (L) + C_2 \,,\,   \overline{V_{\min}^k}^c = C_k \,,\, k > 2 \,;,
C_k = \left. \frac{\dif^k}{\dif t^k} {\ln \left[\frac{\Gamma(1 + t)}{G(1 + t)} \right]} \right\vert_{t=0} \,,
\end{equation}
the first values are $$ \set{C_2, C_3, C_4} = \set{1+\gamma +\frac{\pi ^2}{6},-2\zeta(3) -\frac{\pi ^2}{3},6 \zeta (3)+\frac{\pi ^4}{15}} = \set{3.22215, -5.69398, 13.7063} \,, $$
where $\gamma$ is the Euler constant. {Higher cumulants $C_k$ are also easily expressed in terms of poly-gamma functions,
using the formula Eq. \eqref{phik} in the Appendix.}
The above predictions were already obtained in Ref.~\cite{fyodorov2009statistical} (section 5) {and justified there -- to some extent -- using the fact that the model can be obtained as some limit of an interval model with edge charges (with no pathology before taking the limit). Here we revisit the problem, with a different limiting/cutoff procedure, and make some further clarifying remarks.}

First, we notice that the free energy (minimum) distribution given by eq.~\eqref{eq:mehta} [eq. \eqref{eq:vminmehta}, respectively] has a extensive variance $\propto \ln L$, similarly to that of fBm0 models (pinned logREMs). Yet, the large variance has a different origin: it arises from the fluctuations of the log-correlated field of typical wave length $\sim L$.
%For this reason, the variance is independent of the short-distance cut off $\epsilon$.
Second, we consider the convexity of the cumulant generating function $\kappa(t) := \ln \overline{\exp(t \mathcal{F}_G)}$. It is known that $\ln \Barnes_b (x) \sim x^2 \ln (x^2)$ when $x \to  + \infty$ (see e.g. Ref.~\cite{cao17thesis}, section 2.3.3), so for any fixed $L < \infty$,  by eq.~\eqref{eq:mehta},  $\kappa(t) \sim -t^2 \ln (t^2)$ for large enough $t \gg \ln L$ (it can be shown similarly that the same problem exists with the free energy at any finite temperature). Therefore, adding a fixed cut-off $L$ cannot cure the non-convexity problem for $t \in [0, \infty)$.

We believe that this problem reflects only the non-commutation of the limits $L\to+ \infty$ and $t\to+ \infty$
and as such does not discredit the results above.
For any fixed $t>0$, eq.~\eqref{eq:mehta} becomes exact in the $L\to\infty$ limit. In contrast, for any fixed $L$, eq.~\eqref{eq:mehta} must break down for some large enough $t$ (since the Mehta integral is on the infinite axis), and be replaced by some unknown finite-size expression which should be everywhere convex.

In summary, like fBm0, the free energy distribution of the Gaussian model is  the convolution of a Gaussian with extensive variance and an $O(1)$ correction, which was calculated correctly by the methods of Ref.~\cite{fyodorov2009statistical,fyodorov2010freezing}, despite apparent pathologies, {which, in the present case are due to non-commutativity of limits.}

\subsection{The two-dimensional logREM} \label{sec:DF}
The 2D logREM studied here is defined by a random potential $V(z)$ which is the sum of
a 2d GFF on the complex plane $\phi(z)$, and of a deterministic background potential $U(z)$:
\begin{align} V(z) = \phi(z) + U(z) \,,\,
 U(z) = 4a_1 \ln \abs{z/L} + 4a_2 \ln \abs{(z-1)/L} \label{eq:Uz} \,,\, a_1, a_2 < Q/2 \,,\, a_1 + a_2 > Q/2 \,.  \end{align}
$U(z)$ is characterized by two parameters $a_1, a_2$ (called the \textit{charges}), which should satisfy the above restrictions, see below. This model is the simplest exactly solved 2D logREM (see Ref.~\cite{cao16liouville} for generalization to other geometries).

Introducing again the large scale cut off $L$, {the domain size}, and $\epsilon$, {the short-distance cutoff}, the continuum partition function is written as
\begin{equation}\label{eq:ZDF}
\mathcal{Z}_{\text{2D}} = \int_{\Omega(L)} \frac{\dif^2 z}{\epsilon^2}  e^{-\beta \phi(z)} \abs{\frac{L}{z}}^{4a_1\beta} \abs{\frac{L}{z-1}}^{4a_2\beta} \,,\,  \overline{\phi(z)} =0 \,,\, \overline{\phi(z)^2} = 4 \ln (L/\epsilon) \,,\, \overline{\phi(z) \phi(z')} = 4 \ln \abs{\frac{L}{z-z'}} \,,\, z \neq z' \,.
\end{equation}
Here {the integral domain} is $\Omega(L) = \set{z: \abs{\Re(z)} < L/2, \abs{\Im(z)}<L/2}$, and $\dif ^2 z = \dif x \dif y$. The restriction on the charges in eq.~\eqref{eq:Uz} ensures that the associated Gibbs measure, as $\epsilon \to 0, L \to \infty$, tends to a non-trivial limit which is neither a delta peak at $0$ or $1$, nor zero {everywhere. The disorder-averaged} Gibbs probability density at any point $z$ can be calculated {as} a 4-point correlation function of Liouville field theory~\cite{cao16liouville}. Note that for convenience
%for the sake of an argument below,
we added a constant $\propto \ln L$ to $U(z)$ compared to \textit{op. cit.}, so that $U(z) < 0$ everywhere in $\Omega(L)$.

While Ref.~\cite{cao16liouville} focused on the (well defined) Gibbs measure, here we shall be interested in the free energy fluctuations. Similarly to the Gaussian model, the long wave length fluctuations of the 2d GFF result in a free energy distribution of variance $\sim 4 \ln L + O(1)$. To obtain the non-Gaussian corrections, we consider the replicated partition function, which becomes (as $\epsilon \to 0, L \to \infty$) a Dotsenko-Fateev (DF) integral~\cite{dotsenko1984conformal} when averaged over disorder. The analytical continuation of the DF integral leads to essentially the Dorn-Otto and Zamolodchikov brothers' (DOZZ) structure constant of Liouville field theory~\cite{dorn1994two,zamolodchikov1996conformal} (see Ref.~\cite{kupiainen2017integrability,david2016liouville} for recent rigorous developments). Indeed, assuming unbroken replica symmetry, we find (a self-contained derivation is provided in the Appendix~\ref{sec:dozz})
%\red{should be $L^{+2 t^2}$ no?} \uv{corrected}
\begin{equation}
\overline{\exp(t \mathcal{F}_{\text{2D}})} \,  \Gamma(1 + t/\beta)= \epsilon^{2Qt} L^{-4(a_1 + a_2)t {+} 2 t^2} \,  t\, \DOZZ_b(a_1, a_2,  Q-a_1-a_2+ t )  \,, \label{eq:etfDozz}
\end{equation}
where $\DOZZ_b(a_1,a_2,a_3)$ is the DOZZ structure constant~\footnote{The relation between $\DOZZ$ and $\tDOZZ$ is the same as that between $\mathbf{M}$ and $\tilde{\mathbf{M}}$ in section~\ref{sec:cir}, eq.~\eqref{eq:morrisgamma}. The differing factor will only contribute to the first cumulant, see also below eq.~\eqref{eq:etf2Dlow}. }: %\red{Pierre needs to recheck all formula}
\begin{align}
& \DOZZ_b(a_1,a_2,a_3) = \left[ \gamma(b^2) \pi b^{2-2b^2}  \right]^{(Q-a_1-a_2-a_3)/b} \tDOZZ_b(a_1,a_2,a_3) \,,\, \\
& \tDOZZ_b(a_1,a_2,a_3) := \frac{ \Upsilon'_b(0) }{\Upsilon_b\left(\sum_{j=1}^3 a_j-Q \right)} \prod_{k=1}^{3}\frac{\Upsilon_b(2 a_k)}{\Upsilon_b\left(\sum_{j=1}^3 a_j -2a_k\right)} \,. \label{eq:DOZZ}
\end{align}

Here, $\gamma(x) = \Gamma(x)/\Gamma(1-x)$, and $\Upsilon_b(x)$ is the Upsilon function, which can be related to $\Barnes_b$ by Ref.~\cite{fateev2000boundary}, eq. 3.16,  and \cite{fyodorov2015moments}, below eq. 237, see also Ref.~\cite{cao16liouville}, eq. (C5-C9)}.
\begin{equation}
\Upsilon_b(x) = G_b(x) G_{b}(Q-x) \,,\, G_b(x) = \Barnes_b (x) b^{x (Q-x) / 2} (2\pi)^{x(1-1/b)/2} \,   \,.  \label{eq:UBer}
\end{equation}
$\Upsilon_b(x)$ satisfies functional relations similar to eq.~\eqref{eq:functional}. Both $\Upsilon_b(x)$ and $G_b(x)$ are invariant under the change of variable $b \to 1/b$, i.e., they enjoy the ``duality invariance'' property,  as does the structure constant: $\tDOZZ_b(a_1, a_2, a_3) = \tDOZZ_{1/b}(a_1, a_2, a_3)$.
We recall that in the above equations,
\begin{equation}
Q = b + b^{-1} \,,\, b = \beta \,,\, \beta < 1 \,.
\end{equation}
In the $\beta > 1$ phase, by the {RSB/duality-freezing scenario~\cite{fyodorov2009statistical,fyodorov2015moments}, the LHS of eq.~\eqref{eq:etfDozz} freezes at $b = 1$ ($Q=2$)}, so that eq.~\eqref{eq:UBer} simplifies to a product of two ordinary Barnes functions:
\begin{equation}
\Upsilon_1(x) = G(x) G(2-x) \,.
\end{equation}
 Therefore the $\beta>1$-phase version of eq.~\eqref{eq:etfDozz} is simplified to the following (note that $G(0) = 0$, $G'(0) = 1$ and $G(2) = 1$, so $ \Upsilon'_1(0) = 1$):
\begin{equation}
\overline{\exp(t \mathcal{F}_{\text{2D}})} \,  \Gamma(1 + t/\beta) =\epsilon^{2Qt} { L^{-4(a_1 + a_2)t + 2 t^2}}
\frac{e^{c_1 t} t}{G(t) G(2 - t)}
\prod_{k=1}^3 \frac{G(2a_k)G(2 - 2a_k)}{G(2 a_k - t) G(2 - 2 a_k + t)} \,,\, a_3 := 2-a_1-a_2 + t \,. \label{eq:etf2Dlow}
\end{equation}
where $c_1 = c_1(\beta)$ depends on short-distance details and only affects the first cumulant~\cite{cao16order}. The above equation holds down to the zero temperature limit $\beta\to\infty$. In that limit, the free energy becomes the minimum $V_{\min}$ of $V(z)$ eq.~\eqref{eq:Uz}, for which we predict the non-Gaussian cumulant corrections:
\begin{align}
&\overline{V_{\min}^2}^c =  2 \ln (L^2) + C_2 \,,\,   \overline{V_{\min}^k}^c = C_k \,,\, k > 2 \,;, \label{eq:2Dcorrections} \\
&C_k (a_1, a_2) = \left. \frac{\dif^k}{\dif t^k} \ln \left[ \frac{t}{G(t) G(2 - t)}
\prod_{k=1}^3 \frac{G(2a_k)G(2 - 2a_k)}{G(2 a_k - t) G(2 - 2 a_k + t)} \right] \right\vert_{t=0} \,,\, a_3 := 2-a_1-a_2 + t \,. \label{eq:Ck2D}
\end{align}
Note that the factor $\Gamma(1+t/\beta)$ in eq.~\eqref{eq:etf2Dlow} tends to $1$ when $\beta \to \infty$ and implies eq.~\eqref{eq:Cbetaklow} for low-temperature free energy cumulant corrections. Using eq.~\eqref{eq:Barnes} the cumulants are readily expressed in terms of poly-gamma functions. The  explicit expression of cumulants $C_k(a_1,a_2)$ is given in \eqref{Ck2D} in Appendix~\ref{sec:dozz}
where their dependence in the charges $a_1,a_2$ is studied in some details.

We now turn to the analytical properties of the moment generation function eq.~\eqref{eq:etfDozz}, which we would expect to resemble  that of the above discussed Gaussian model, given that they are both defined on unbounded domains. However,  while eq. \eqref{eq:mehta} never vanishes for $t > 0$, eq. \eqref{eq:DOZZ} has the first zero at $t \to a_1 +  a_2 - Q/2$, which comes  from the Upsilon factor $\Upsilon_b(2 a_3) = \Upsilon_b(2(Q-a_1-a_2 + t))$ [by eqs. \eqref{eq:UBer} and \eqref{eq:Barneszeros}, $\Upsilon_b(x)$ has a zero at $x = Q$].

To reveal the physical significance of this zero, we come back to the leading behavior of the free energy $\mathcal{F}_{\text{2D}}$, which, according to eq. \eqref{eq:etfleading}, has a Gaussian distribution of mean value $2 Q \ln \epsilon - 4(a_1 + a_2) \ln L$ and variance $4 \ln L$, up to $O(1)$-corrections, whose cumulants can be obtained from eq. \eqref{eq:etfDozz}. Yet, we argue now that the {Gaussian tail of the free energy distribution} cannot prevail in the whole large deviation regime. To see this, let us ignore the restrictions of eq.~\eqref{eq:Uz}, and consider the 2d logREM with $a_1 = a_2 = 0$, that is, \textit{without} the background potential. Then we have an ordinary logREM with size $M = (L/\epsilon)^2$ and by eq.~\eqref{eq:freezing}, its free energy $\mathcal{F}_{\phi}$ has mean value $2 Q \ln (\epsilon/L) + o(\ln(\epsilon/L))$ and variance of order unity. Now,  since $U(z) < 0$ for any $z \in \Omega(L)$, {$\mathcal{F}_{\text{2D}}<\mathcal{F}_{\phi}$} for any fixed 2d GFF realization $\phi(z)$. We deduce that the Gaussian tail of cannot continue beyond $\sim 2Q \ln (\epsilon/L)$; as a result, the large deviation function of $\mathcal{F}_{\text{2D}}$ (using $2\ln L$ as the large variable, with $\epsilon$ small but fixed) has a hard wall:
\begin{equation}  \mathcal{L}(f_{\text{2D}}) := - \frac{1}{2 \ln L} \ln \mathrm{Prob.}(\mathcal{F}_{\text{2D}} = f_{\text{2D}} \ln (L^2) + o(\ln L)) =
\begin{cases} (f_{\text{2D}} + 2(a_1 + a_2))^2 / 4  &  f_{\text{2D}} < -Q \\ +\infty  &  f_{\text{2D}} \geq -Q   \end{cases} \,, \label{eq:ldfDF}   \end{equation}
Note that this large deviation function has the form of a Gaussian cut off by a hard wall, reminiscent of the fBm0 case, eq. \eqref{eq:ldf}. Therefore, the moment generating function $\overline{\exp(t \mathcal{F}_{\text{2D}})}$ has also a termination point transition, which occurs precisely at $t = a_1 + a_2 - Q/2$ according to eq.~\eqref{eq:ldfDF}. So eq. \eqref{eq:etfDozz} should be amended in the following way:
\begin{equation} \ln \overline{\exp(t \mathcal{F}_{\text{2D}})} \sim \begin{cases}  Qt \ln(\epsilon^2) + (t^2 -2(a_1 + a_2)t) \ln(L^2)  & 0 \leq t < a_1 + a_2 - Q/2 \\
  Qt \ln(\epsilon^2) - Qt \ln (L^2)   & t \geq a_1 + a_2 - Q/2
 \end{cases} \label{eq:etF2dterm}  \end{equation}
where ``$\sim$'' refers to equality modulo $o(\abs{\ln \epsilon}), o(\ln L)$ terms. The physical origin of the zero at $t = a_1 + a_2 - Q/2$ is now revealed. It is of the same nature as the ``problematic'' zero of fBm0 models: it signals another termination point transition and the associated log-corrections.

We close  this section with some comparing remarks. First, the 2D logREM has such a termination point transition thanks to the logarithmic growth of the background potential $U(z)$. In contrast, the Gaussian model has a quadratic background potential, so it has no termination point transition (hence no problematic zero).

 Second, the 2D logREM's termination point transition is of long-distance nature (the large number is $L^2$, not the number of sites $M = (L/\epsilon)^2$), while that of fBm0 is of short-distance nature (the large number is $M = 1/\epsilon$). Nevertheless, it is interesting to notice that, in terms of $a_3 = Q-a_1-a_2 +t$, the third charge (``charge at infinity'') appearing in the DOZZ structure constant~\eqref{eq:etfDozz}, the termination phase of eq.~\eqref{eq:etF2dterm} is described as $a_3 > Q/2$. This inequality is known in Liouville field theory as the violation of the Seiberg bound, which is associated to short-distance termination point transitions, in geometries where the charge is not at infinity~\cite{cao16liouville,cao17seiberg}. Thanks to the conformal invariance of Liouville field theory, the short- and large-distance transitions are nicely unified.

% \footnote{The short-distance critical behaviors of the 2D logREM, besides freezing, occur around each of the charges at $z=0, 1$, as studied in Ref.~\cite{cao17seiberg}; when $a_1, a_2  < Q/2$, it turns out that there can be only first-order transition in the \textit{negative} free energy large deviation regime.}.

%\red{Maybe you could mention also that if one uses conformal invariance (or go to sphere with 3 charges)
%	these seemingly different transitions (from small scale or large scale) become essentially the
%	same - do you agree with that?}
%
%\uv{Thanks for the nice suggestion! Raoul had  remarks in the same direction regarding that long paper. I added a lightweight discussion, and I also minimized references to that paper in preparation  in this section, to avoid overloading the reader with LFT. }

Finally, the large deviation function eq.~\eqref{eq:ldfDF} is identical to that of logREMs with one charge $a = a_1 + a_2 > Q/2$~\cite{cao17seiberg}. This is understandable, since the two charges of the DF model merge into one seen, viewed from the scale $L$. From this viewpoint, the phase $f_{\text{2D}}<-Q$ can be called a \textit{bound} phase, in the sense that the thermal particle is confined in a region of size $\sim 1$, much smaller than the system size $L$, whereas the phase where $f_{\text{2D}} = -Q$ is associated to rare realizations just like the termination/pre-freezing phase. (indeed they can be unified as the \textit{critical} phase  in a broader framework~\cite{cao17seiberg}).
 Remarkably, thanks to the presence of two charges at the intermediate $O(1)$ scale, there is a non-trivial Coulomb gas integral in the bound phase, providing an {integrable} signature of the bound-critical transition that was not possible in simple one-charge models considered in Ref.~\cite{cao17seiberg}.

\section{Conclusion}
In this work {we suggested a cure} to the {apparent} pathologies that plagued some logREMs associated with exact solvable Coulomb gas integrals: those defined on unbound regions {and fBm0 models, which can be seen as pinned logREMs.} The common origin of the {``problems''} turns out to be the extensive variance of the free energy distribution. Recognizing the importance of this feature allows benign reinterpretation of the apparent pathologies in the calculation of the cumulants.
As a result we give here non trivial predictions (some tested numerically) for the cumulants of the distributions of the free energy and the global minimum value for the fBm0 models (bridge and interval) and for the 2D logREM.
Furthermore, the ``problematic zero'' of the free energy moment generating function is not a signal of the breakdown of the method used, but rather is a signature of a termination point/prefreezing transition and of the associated emergence of log-corrections.

Nevertheless, some issues still call for a deeper understanding. In particular the question of the convexity of the moment generating function, especially in presence of the termination point transition. Since the convexity fails \textit{before} hitting the termination-point zero, a satisfactory discussion of this point would require mastering the \textit{finite-size} properties of the termination point transition. From a broader perspective, finite-size properties of glassy transitions in general log-correlated REMs are by themselves a hard but valuable problem for future study.

 \vspace{.2cm}
 \textit{Acknowledgments}.  We acknowledge D. Ostrovski for pointing out first the importance of extensive free energy variance.  We thank A. Rosso and R. Santachiara for discussions and collaborations on related projects. X.C acknowledges support from a Simons Investigatorship, Capital Fund Management Paris, and Laboratoire de Physique Th\'eorique et Mod\`eles Statistiques. PLD acknowledges support from ANR grant ANR-17-CE30-0027-01 RaMaTraF. The research at King's college was supported by EPSRC grant EP/N009436/1, ``The many faces of random characteristic polynomials''.

 \appendix
 \section{ Log-corrections induced by the termination point/prefreezing transition}\label{sec:free2}

 \begin{figure}
 	\begin{center}
 		\includegraphics[scale=.45]{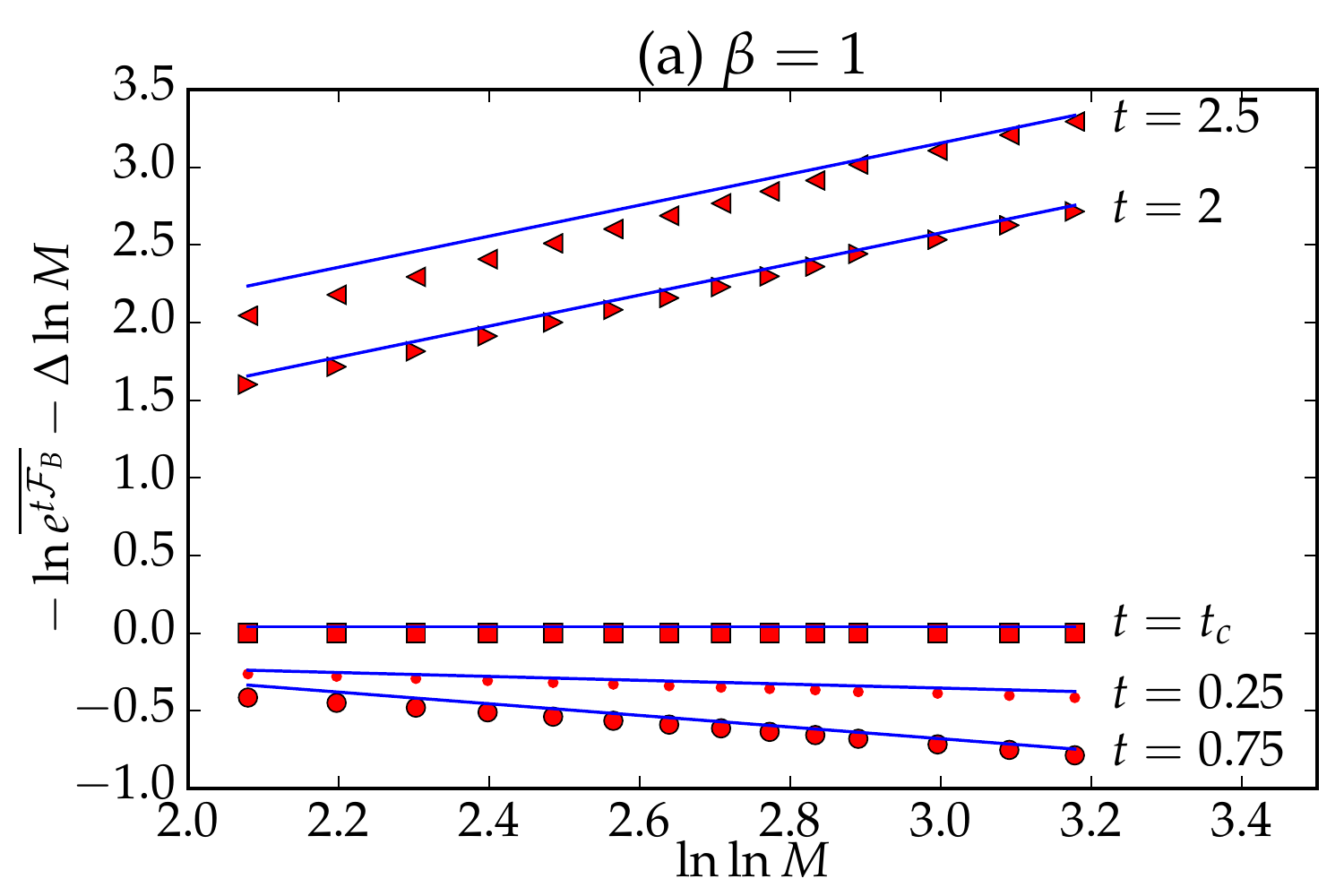}
 		\includegraphics[scale=.45]{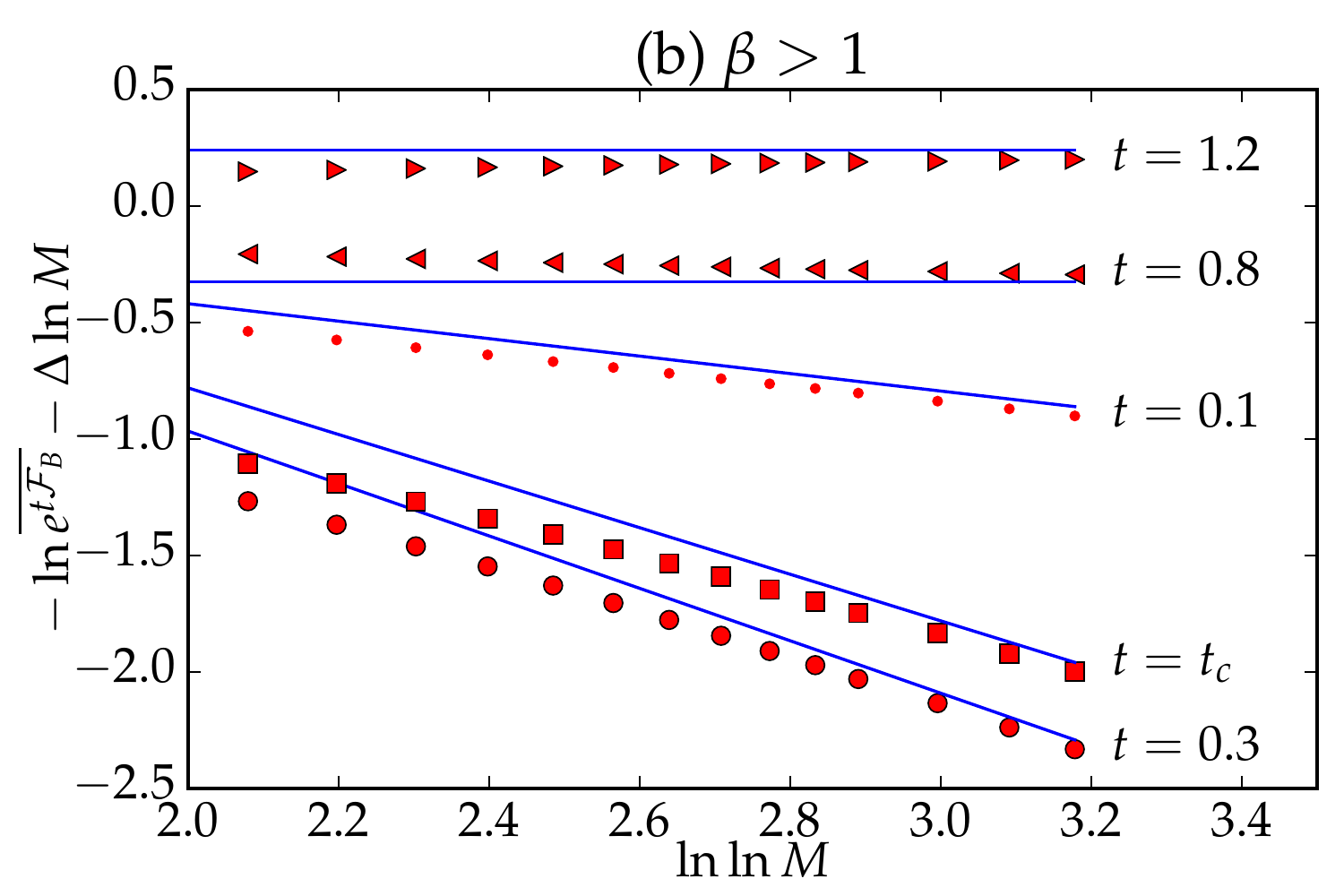}
 	\end{center}
 	\caption{Numerical test of the predictions of the termination point transition, at the freezing transition $\beta = 1$ (a) and in the frozen $\beta > 1$ phase (b, $\beta = 2.5$). The blue lines represent the log-correction predictions in eq. \eqref{eq:logfreeze1} and \eqref{eq:logfreeze2}. Red markers represent the numerical data obtained in the circular model (see~\cite{fyodorov2009statistical} for simulation method), with sizes $M = 2^8, \dots, 2^{24}$ (Translation invariance is used to enhance the statistics), with various values of $t$, above and below the critical value $t_c = Q/2$. The leading behaviour  $\Delta \ln M$ is extracted, according to eq. \eqref{eq:etfleading}. The analogue test for $\beta < 1$ can be found in Ref.~\cite{cao17thesis}, p114.
 	}\label{fig:termination}
 \end{figure}
{Recently~\cite{cao16liouville,cao17thesis} a relation between logREMs and Livouille field theory has been exploited to predict the new subleading logarithmic factors to the free-energy moment generating function.
Such corrections  are reminiscent of those arising at the point of the freezing transition [eq. \eqref{eq:freezing}], but arise instead  in the high-temperature phase $\beta<1$ at the termination point/prefreezing transition. Namely, when $\beta<1$ one needs to replace the leading order  expression eq.~\eqref{eq:etfleading} with a more accurate expression
 \begin{equation}
 \overline{e^{t\mathcal{F}_B}}  = \overline{p_{\beta,1}^{t/\beta}} =\begin{dcases}
 M^{-t Q + t^2} \times O(1) &  t < Q/2 \\
 M^{-Q^2/4} \left(\ln M\right)^{-\frac12} \times O(1) & t = Q/2 \\
 M^{-Q^2/4} \left(\ln M\right)^{-\frac32} \times O(1) & t > Q/2
 \end{dcases} \,,\,\,\,\quad  \beta < 1 \,. \label{eq:termination}
 \end{equation}

In the $\beta > 1$ phase (and at the freezing transition $\beta = 1$), the log-corrections are however different, and remained so far inaccessible by the mapping to Liouville field theory~\cite{cao16liouville,cao17thesis}. Here we aim to filling in this gap by means of a simple argument, supported by a numerical study.

  When $\beta > 1$  the termination point transition (see  eq. \eqref{eq:freezing}) happens at $t = Q / 2 = 1$. Although for $t < 1$ the system is not in the termination point-dominated phase,  there is still a log-correction to $\mathcal{F}_B = \mathcal{F} - V_1$ stemming from that of the ordinary logREM free energy $\mathcal{F}$. Our idea is to establish the log-correction to $\mathcal{F}_B$ by neglecting correlations between the free energy $\mathcal{F}$ and the value $V_1$ of the potential in the ordinary logREM:
 \begin{equation}
 \overline{e^{t\mathcal{F}_B}} =    \overline{e^{t\mathcal{F} - t V_1} }  \approxeq \overline{e^{t\mathcal{F}}} \times \overline{\exp(- t V_1)} \approxeq e^{t\overline{\mathcal{F}}} \times \exp\left(\frac{t^2}{2}  \overline{V_1^2}\right)=
 M^{-2t + t^2} \left( \ln M \right)^{\chi t}  \times O(1) \,,\, t < 1 \,,\, \chi = \begin{cases} \frac12 & \beta= 1 \\ \frac32 & \beta > 1 \end{cases}  \,, \label{eq:logfreeze1}
 \end{equation}
  where we have used eq. \eqref{eq:freezing} and eq. \eqref{eq:circledef}.  Note that, by setting $t \ll 1$, the above  equation reduces to eq. \eqref{eq:freezing}. For finite $t$, we find that eq. \eqref{eq:logfreeze1} is consistent with the results of our numerical simulations, see Figure \ref{fig:termination}. The factorization approximation in eq.~\eqref{eq:logfreeze1} can be justified with the following, heuristic but plausible, argument: since the free energy $\mathcal{F}$ is dominated by the deepest minima of the potential $V$, it is strongly correlated with $V_1$ only in such realizations when $V_1$ is close to one of such minima.  As the deepest minima happen randomly in of the order of one sites in the sample, the correlations in question happen with probability of order of $1/M $ which for $t < 1$ is much smaller than $\ll M^{-2t + t^2}$. Hence the error made by omitting contributions from such events should be sub-dominant compared to eq.~\eqref{eq:logfreeze1}.

  When $t \geq Q/2 = 1$, the quantity $\overline{e^{t\mathcal{F}_B}} $ becomes dominated by realizations in which the free energy receives significant contribution from the site $j=1$, which is close to a deep minimum. Heuristically, we may expect that when $t$ crosses the  value $Q/2$ in the $\beta>1$ phase the change in the log-corrections  should be identical to what happens in the $\beta < 1$ phase, since it should be determined by the potential structure around the site $j=1$. This mechanism is suggestive of predicting the following behaviour:
 \begin{equation}
 \overline{e^{t\mathcal{F}_B}} = \begin{cases}
 M^{-1} \left(\ln M\right)^{\chi-\frac12} \times O(1) & t = 1 \\
 M^{-1} \left(\ln M\right)^{\chi-\frac32} \times O(1) & t > 1
 \end{cases} \,,\, \beta \geq 1 \,, \label{eq:logfreeze2}
 \end{equation}
 where $\chi=1/2$ for $\beta=1$ and $\chi=3/2$ for $\beta>1$, i.e. the same as given in eq. \eqref{eq:logfreeze1}.  We find that eq. \eqref{eq:logfreeze2} is again in a nice agreement with the numerical data, see Figure~\ref{fig:termination}. We perform some further consistency checks. First, let us consider a special case $t = \beta$. Then for $\beta<1$ we have $t=\beta<Q/2=\frac{1}{2}(\beta+\beta^{-1})$, hence eq. \eqref{eq:termination} shows that the log-correction is absent.
  They are also absent for $t=\beta$ as long as $\beta\ge 1$, as follows from eq. \eqref{eq:logfreeze2}. This should be no surprise, since for $t=\beta$  eq. \eqref{eq:ZBisp} implies $e^{t\mathcal{F}_B} = p_{\beta,1}$ which is simply the Gibbs probability weight of an ordinary logREM. Restricting further to translation invariance systems (such as the circular model), we then have the identity $1 = \sum_{j=1}^M \overline{p_{\beta,j}} = M \overline{p_{\beta,1}}$, thus $\overline{p_{\beta,1}}= M^{-1}$ with no approximation, leaving no possibility of log-corrections.  For general values of $t>0$ and $\beta > 1$, we observe that eqs.~\eqref{eq:logfreeze1} and \eqref{eq:logfreeze2} imply that $\overline{p_{\beta,1}^t} \gtrsim 1/M \times O(1)$.  This is consistent with the presence of a few ( i.e. of order unity) Gibbs weights $p_{\beta,j} \sim O(1)$ in typical samples in the low-temperature phase: indeed, a consequence is that the site $j = 1$ has such a Gibbs weight with probability $\sim 1/M$, leading to $\overline{p_{\beta,1}^t} \gtrsim 1/M \times O(1)$.

  Finally, we remark  that for the uncorrelated REM the log-correction is also absent in the  $\beta  > 1$, $t  > 1$ phase, as was shown in Ref.~\cite{fyodorov2009pre}. Indeed, eq. (9) in that paper implies
 \begin{equation}
 \overline{e^{t\mathcal{F}_B^{\text{REM}}}}  \to  \frac{1}{M} \frac{\Gamma((t-1)/ \beta)}{\Gamma(t/\beta) \Gamma(1-1/\beta)} \,,\, \quad t > 1, \beta > 1 \,, \label{eq:FBREMa}\end{equation}
 where $\mathcal{F}_B^{\text{REM}}$ is given by eq. \eqref{eq:FBmulti}, but for uncorrelated REM. We see that when $t \to 1_+$ (with $\beta>1$ fixed), the $\Gamma$-factor in the numerator diverges to $+\infty$, suggesting a log-correction in the regime $t < 1, \beta > 1$. Indeed, below we derive the $t<1$ counterpart of eq.(\ref{eq:FBREMa}) :
\begin{equation}  \overline{e^{t\mathcal{F}_B^{\text{REM}}}} \approx M^{t^2 - 2t} \left(\ln M\right)^{\frac12 t}  \frac{(4\pi)^{t/2}}{\Gamma(1-1/\beta)^t}\frac{\Gamma(1  + t)}{\Gamma(1 + t/\beta)} \,, \quad t < 1 \,,\, \beta > 1 \,.  \label{eq:FBREM} \end{equation}
 as $M \to \infty$. Thus, a log-correction appears also in the standard REM case, albeit with an exponent different from eq.~\eqref{eq:logfreeze2}. This is not at all surprising as it is well-known that the free energy of the REM has a $\frac12 \ln \ln M$ correction in the frozen phase, rather than $\frac32 \ln \ln M$ typical for log-REMs.

{ We now derive  eq.~\eqref{eq:FBREM} for the REM}. Let $V_1, \dots, V_M$ be the Gaussian energy levels of the REM, so that $\overline{V_i} =0$ and $\overline{V_i V_j} = \delta_{ij} 2 \ln M$. Let $\mathcal{Z} = \sum_{j=1}^M e^{- \beta V_j}$ be the REM partition function, then $\mathcal{F}_B^{\text{REM}} := \mathcal{F}_B = - V_1 - \beta^{-1} \ln Z$ as eq.~\eqref{eq:FBmulti} (here and below we drop the "REM" subscript).

The ensemble average featuring in the left-hand side of eq.~\eqref{eq:FBREM} can be re-written as the following integral:
\begin{equation} \label{eq:FBREMint}
\overline{e^{t\mathcal{F}_B}}  = \overline{  e^{- tV_1} \mathcal{Z}^{-t/\beta}} = \frac{\beta}{\Gamma(t/\beta)}   \int_{-\infty}^{+\infty} \overline{ e^{- t(V_1-y)}  \exp(- e^{\beta y} \mathcal{Z})   } \dif y  \,.
\end{equation}
Now we calculate the integrand in the right-hand side. Writing $\exp(- e^{\beta y} \mathcal{Z}) = \prod_{i=1}^M \exp(- e^{\beta (y-V_i)}) $ and exploiting that  $V_i$'s are independent and identically distributed, we have
\begin{equation} \overline{e^{- t(V_1-y)}  \exp(- e^{\beta y} \mathcal{Z})} = g(t,y)\left[ g(0,y)\right]^{M-1} \,,\, g(t,y) := \overline{e^{- t(V_1-y)} \exp\left(-e^{\beta(y-V_1)} \right)  } \,. \label{eq:REMfactor}  \end{equation}
Note that function $g(0,y)$ is identical to one denoted $\gamma(y)$ and computed in the last appendix of Ref.~\cite{cao16order} (see also \cite{cao17thesis}, 2.17-2.22): when $M\to\infty$, $g(0,y)^{M-1}$  tends to a Gumbel double-exponential shifted by the extensive free energy of REM; more precisely, we have:
\begin{equation}
\left[g(t,\delta + F)\right]^{M-1} \stackrel{M\to\infty}{\longrightarrow} \exp\left( - e^{\delta} \right) \,,\,  F := - 2\ln M + \frac12 \ln \ln M - \ln \Gamma(1-1/\beta) + \ln (\sqrt{4\pi}) \,,\, \beta > 1 \,. \label{eq:REMgamma}
\end{equation}
The method suggested in Ref.~\cite{cao16order} and based on a variant of the Hubbard-Stratonovich transformation readily generalizes to any $t > 0$. Employing it, one has the following intergral representation:
\begin{equation}
g(t,y) = \int_{-t+\epsilon+\im \R} \frac{\dif p}{2 \pi \im} e^{p^2 \ln M - p y} \beta^{-1} \Gamma((t + p)/\beta) \,,
\end{equation}
where the  contour of integration runs in the complex plane parallel to the imaginary axis, with a fixed real part $>-t$.  We then evaluate the above integral in the saddle-point approximation, justified by $\ln M\gg 1$. The saddle point is at $p_* = y / (2\ln M)$. Now, when $t < 1$, $p_* < -t$ for any $y / \ln M \in (-\infty,-2t)$, and to deform the contour through the saddle point one has to cross a pole of the Gamma-function at $p = -t$, which gives the dominant contribution to the integral:
\begin{equation} g(t,y)  = e^{t^2 \ln M + t y} + \text{subleading terms} \,,\, y / \ln M \leq -2t \,. \label{eq:gty} \end{equation}
When $  y / \ln M > -2t > -2$, the value of the factor $\left[g(0,y)\right]^{M-1}$ given by  eq.~\eqref{eq:REMgamma} is so small that the precise behavior of the factor $ g(t,y) $ is immaterial for the value of the integral eq.~\eqref{eq:FBREMint}.  Combining expressions eq.~\eqref{eq:FBREMint} to \eqref{eq:gty} we see that
\begin{equation}
\overline{e^{t\mathcal{F}_B}} \approx \frac{\beta}{\Gamma(t/\beta)}  \int_{-\infty}^{\infty}   e^{t^2 \ln M + t (\delta + F)}   \exp\left( - e^{\delta} \right)  \dif \delta
= M^{t^2-2t} (\ln M)^{\frac12 t}  \frac{(4\pi)^{t/2}}{\Gamma(1-1/\beta)^t}  \frac{\Gamma(1+t) }{\Gamma(1 + t/\beta)}  \,,
\end{equation}
which is eq.~\eqref{eq:FBREM}. We remark that the log-correction can be traced precisely to that of the REM free energy in the $\beta > 1$ phase, see eq.~\eqref{eq:REMgamma}.}

\section{Replica approach to 2D logREM and Dotsenko-Fateev integrals}\label{sec:dozz}
{Here we outline} the steps that lead to eq.~\eqref{eq:etfDozz}, in the most self-contained way possible, and without assuming knowledge from Liouville field theory. Using the replica-trick and assuming replica symmetry (in particular, recall $b = \beta$ and $Q = b + b^{-1}$), we calculate the integer moments of $\mathcal{Z}_{\text{2D}}$, eq.~\eqref{eq:ZDF}, in the $L\to\infty$ limit :
%\red{should be
%	$L^{-2 t^2}$ no?} \red{recall definition of $Q$}\uv{yes}
\begin{equation}
\overline{\mathcal{Z}_{\text{2D}}^n}  \epsilon^{-2Qt} L^{4 (a_1 + a_2)t - 2t^2} \stackrel{L\to\infty}{\longrightarrow}
{\int_{\C^n} } \prod_{i=1}^{n} \left[ \abs{z_i}^{-4a_1b }
\abs{1-z_i}^{-4a_2b } \dif^2 z_i \right]  {\prod_{i<j} \abs{z_i - z_j}^{-4b^2}} := \text{DF}(n,b,a_1,a_2)
\end{equation}
The right-hand-side is known as the Dotsenko-Fateev (DF) integral and has the following exact expression~\cite{dotsenko1984conformal} whenever it converges:
\begin{equation}\label{eq:DF}
\text{DF}(n,b,a_1,a_2) =  n!  \frac{\pi^n}{\gamma(-b^2)^n}
\frac{{\prod_{k=1}^n \gamma(-kb^2)}}{ \prod_{j=0}^{n-1} \left[ \gamma(2b a_1 + jb^2) \gamma(2b a_2 + jb^2)
	\gamma(2b a_3 + j b^2) \right]} \,,\, a_3 :=   Q   - a_1 - a_2 - nb \,.
\end{equation}
where $\gamma(x) = \Gamma(x) / \Gamma(1-x)$. In order to analytically continue eq.~\eqref{eq:DF} to $n$ complex, we apply the functional relation eq.~\eqref{eq:functional} to each chain of Gamma functions,
\begin{align}
& \frac{1}{n!}  \text{DF}(n,b,a_1,a_2)=  \lim_{\varepsilon\to0} \left[ (\pi / \gamma(-b^2) )^{n}
\frac{\Barnes_{b}(\varepsilon) \Barnes_{b}(Q)}{\Barnes_{b}(\varepsilon - nb)\Barnes_{b}(Q + nb)}
\prod_{k=1}^3 \frac{\Barnes_{b}(2a_k)\Barnes_{b}(Q - 2a_k)}{\Barnes_{b}(2 a_k + n b) \Barnes_{b}(Q - 2 a_k - n b)} \right] \nonumber  \\
=& \left. \mathrm{Res}_{t\to - nb} \left[
(\gamma(-b^2) / \pi )^{t/b}
\frac{\Barnes'_{b}(0) \Barnes_{b}(Q)}{\Barnes_{b}(t)\Barnes_{b}(Q - t)}
\prod_{k=1}^3 \frac{\Barnes_{b}(2a_k)\Barnes_{b}(Q - 2a_k)}{\Barnes_{b}(2 a_k - t) \Barnes_{b}(Q - 2 a_k + t)}  \right]\right\vert_{a_3 := Q-a_1-a_2 + t} \,.  \label{eq:residue}
\end{align}
In the first line, we introduced an infinitesimal $\varepsilon$ for the product $ \prod_{k=1}^n \Gamma(-kb^2) $ in the numerator of eq.~\eqref{eq:DF}: when $n$ is a positive integer, both $\Barnes_{b}(\varepsilon) $ and $ \Barnes_{b}(\varepsilon - nb) $ tend to $0$ as $\varepsilon \to 0$ but their ratio tends to $ \prod_{k=1}^n \Gamma(-kb^2) $. Then we interpreted that limit as a residue, and re-defined $a_3$ in function of $t = -nb$. Now, observe that any analytical continuation of $\text{2D}(n,b,a_1,a_2) $ to $n$ complex should satisfy the following relation (since $\Gamma(x)$ has a simple pole at $x=-n$ with residue $(-1)^n / n!$, $n = 0, 1,2, \dots$ ):
\begin{equation}    \text{Res}_{t\to - nb} \left[ \Gamma(t/b)  \text{DF}(-t/b,b,a_1,a_2) \right]  = \frac{(-1)^n b}{ n!}  \text{DF}(n,b,a_1,a_2)\,, n = 0,1,2,\dots\,. \end{equation}
Comparing eq.~\eqref{eq:residue} and the above one leads to the following analytical continuation of $\text{2D}(n=-t/b,b,a_1,a_2)$ (this method of analytical continuation is well known in the context of conformal field theory, see e.g.~\cite{zamolodchikov1996conformal}, section 3):
\begin{equation}\label{eq:continuedDF}
\text{DF}(-t/b,b,a_1,a_2) \Gamma(1+t/b)  =    (-\gamma(-b^2) / \pi )^{t/b}
\frac{t \Barnes'_{b}(0) \Barnes_{b}(Q)}{\Barnes_{b}(t)\Barnes_{b}(Q - t)}
\prod_{k=1}^3 \frac{\Barnes_{b}(2a_k)\Barnes_{b}(Q - 2a_k)}{\Barnes_{b}(2 a_k - t) \Barnes_{b}(Q - 2 a_k + t)}  \,.
\end{equation}
As a consistency check, we note that when $t = 0$, the right-hand-side tends to $1$ as does the left-hand-side, since $\Barnes_b$ has a simple zero at $0$.  Simplifying eq.~\eqref{eq:continuedDF} using eq.~\eqref{eq:UBer} and noting $a_3 = Q - a_1 - a_2 + t$, we obtain eq.~\eqref{eq:etfDozz} after some algebra.

Performing the derivatives in eq.~\eqref{eq:Ck2D} we obtain the general result for the cumulants in terms of poly-gamma functions, for $k \geq 2$ as
\begin{eqnarray}
C_k = d_k +  (-1)^{k+1} (\phi_k(2) + \tilde \phi_k(a_1)+ \tilde \phi_k(a_2) )
+  (2^k-1)  \tilde \phi_k(2-a_1-a_2)  \label{Ck2D}
\end{eqnarray}
where we have defined the constants $d_k$ and functions $\tilde \phi_k$ and $\phi_k$
as follows (see also)
\begin{eqnarray}
&& d_k =\left. \frac{d^k}{dt^k} \ln \left(t/G(t)\right) \right|_{t=0}  =
\begin{cases}
 \zeta (2)+\gamma +1 & k=2 \\
 (-1)^k (\zeta (k-1)+\zeta (k)) \Gamma (k) & k\geq 3
\end{cases} \\
&& \phi_k(x)=\frac{d^k}{dt^k} \ln G(x+t)|_{t=0}
= (k-1) \psi ^{(k-2)}(x)+(x-1) \psi ^{(k-1)}(x) - \delta_{k,2} \label{phik} \\
&& \tilde \phi_k(x)=  \phi_k(2 x) + (-1)^k \phi_k(2-2 x)
\end{eqnarray}
Some plots are made from them in Fig.~\eqref{fig:C2D}. Note that $C_k$'s are all symmetric in $a_1$ and $a_2$. Note that the sum of three $\tilde \phi_k$
functions in \eqref{Ck2D} can be loosely interpreted as arising from three independent random variables
associated to each charge $a_j$, $j=1,2,3$, however since the "charge at infinity"
$a_3=2-a_1-a_2$ the last term provides
a coupling between the contributions of charges $a_1$ and $a_2$.} It is important to recall that the charges should be inside the triangular region (see eq.~\eqref{eq:Uz} with $Q=2$ in $\beta>1$ phase) for
the model to be well-defined (see discussion in the text)
%\red{Should not the sum of charges be $2$ rather than $1$? I corrected please check.. also what does $\simeq$ means??} \uv{corrected, changed to =}
\begin{equation}
 \mathcal{R} =  \set{(a_1, a_2)\vert a_1, a_2, < 1, a_1 + a_2 > 1} =  \set{(a_1, a_2,a_3)\vert a_1, a_2, a_3 < 1, a_1 + a_2 + a_3 = {2}} \,.  \label{eq:region}
  \end{equation}
{As a concrete example let us give the special value for the point at the center of the triangle in Fig.~\eqref{fig:C2D} corresponding to charges $a_1=a_2=\frac{3}{4}$.
\begin{eqnarray}
\left\lbrace  C_2  \,,\,  C_3  \,,\,  C_4 \right\rbrace_{a_1 = a_2 = 3/4} = \left\lbrace  \ln(256) \,,\, -32 \zeta(3)  \,,\,  0 \right\rbrace \,.
\end{eqnarray}
} % \uv{agreed on special values above and 0.66, 0.45 cases (against my plotting interface).  commented red and tiny texts. }

\begin{figure}
	\center (a) \includegraphics[height=.2 \textwidth, valign=t]{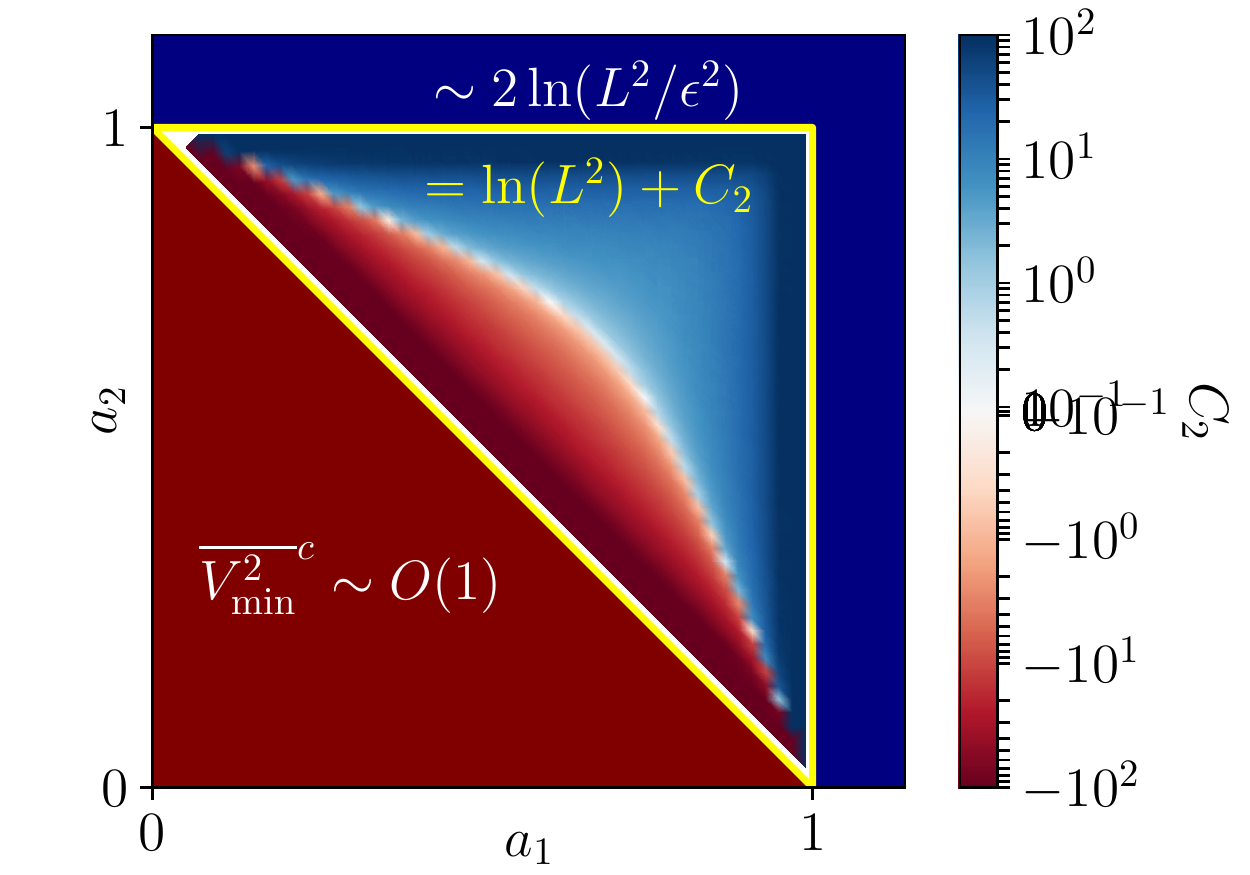}
	(b) \includegraphics[height=.2 \textwidth, valign=t]{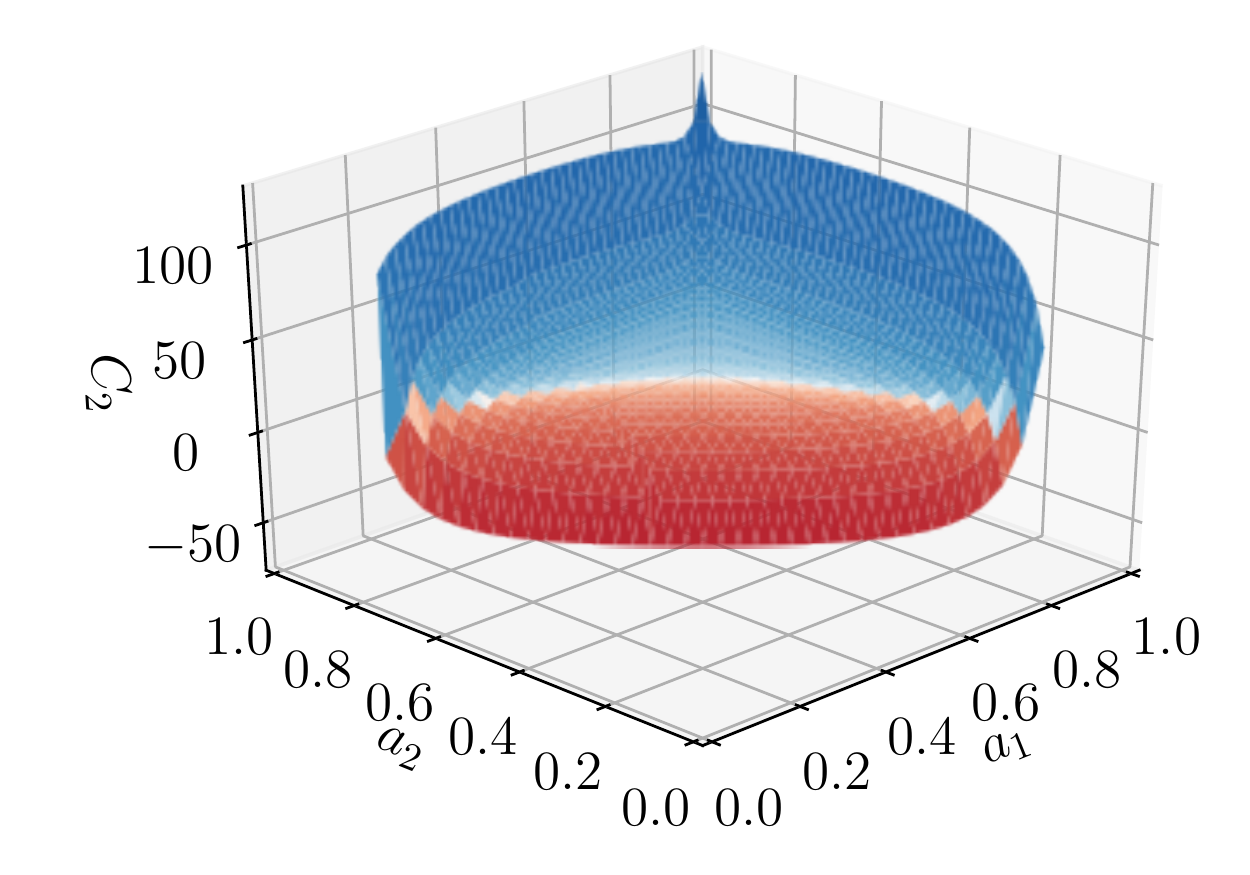}  (c) \includegraphics[height=.2\textwidth, valign=t]{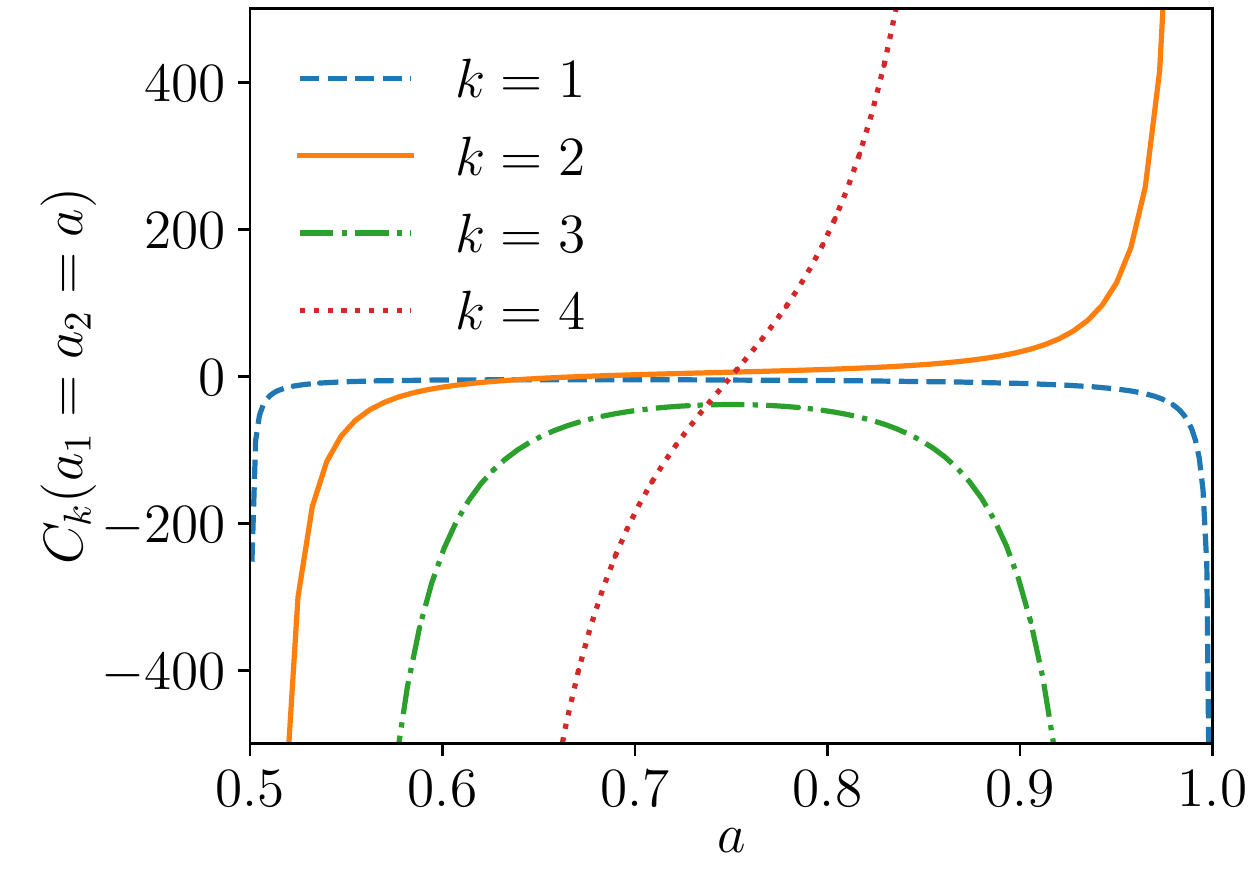}
	\caption{Low-order cumulant corrections to the minimum distribution of 2D logREM, eq.~\eqref{Ck2D}. (a) The variance correction $C_2$ in the whole region $\mathcal{R}$ eq.~\eqref{eq:region} (delimited by the yellow triangle). The behavior of the minimum variance in the short-distance bound phase (blue polygon on top-right) and the large-distance escaping phase (red triangle on bottom-left) the 2D logREM is also indicated [see eqs.~\eqref{eq:bindingvar} and \eqref{eq:escapingvar}]. (b) 3D version of (a) inside the region  $\mathcal{R}$. (c) First cumulants $C_k$ along the line $a_1 = a_2 = a \in (1/2,1)$. $C_1$ is defined in the same way as eq.~\eqref{Ck2D}. $C_{1,3}$ diverge negatively when approaching any boundary of $\mathcal{R}$, while $C_{2,4}$ diverge positively (negatively) when undergoing a binding (unbinding) transition, respectively.} \label{fig:C2D}
\end{figure}
{Since the poly-gamma function $\psi^{(n)}(x)$ is regular everywhere for $x > 0$, with a pole at $x=0$: $\psi^{(n)}(x) = n! (-1)^{n+1}/x^{n+1} + O(1)$, the above formulas eq.~\eqref{Ck2D} diverge rapidly when the parameters approaches the boundary of $\mathcal{R}$. More precisely
\begin{eqnarray}
&& \phi_k(x) = \frac{ (k-1)! (-1)^{k+1}}{x^{k}} + O(1) \quad , \quad
 \tilde \phi_k(x)=  2^{-k} (k-1)!
\left( \frac{(-1)^{k+1}}{x^{k}} -  \frac{1}{(1-x)^{k}} \right) + O(1)
\end{eqnarray}
Hence the singular part of $C_k$ in the allowed domain $\mathcal{R}$ is easily written as
\begin{eqnarray}
&& C_k =  (k-1)!  2^{-k}  \sum_{j=1,2}
\left( \frac{1}{a_j^{k}} + \frac{(-1)^{k}}{(1-a_j)^{k}} \right)
+ (1-2^{-k} ) \left[ \frac{(-1)^{k+1}}{(2-a_1-a_2)^{k}} -  \frac{1}{(a_1+a_2-1)^{k}} \right]
+ O(1)  \label{eq:Cksingular}
\end{eqnarray}
where the $O(1)$ part remains regular when approaching the boundary of $\mathcal{R}$. Formally, near each boundary the cumulant corrections have  two statistically independent pieces: one divergent part scaling as the inverse distance to the boundary, and one of order unity.
%\red{Can you recheck this ? I think one can discuss and interpret the {\it sign} of the
%	divergence at each boundary for each cumulant, can you do it? You do it below so can you then
%	refer to this equation, or maybe merge the discussions. Maybe since divergence
%	is simply of the form $1/x^k$ one cand define/find some scaled (fictitious) PDF at each boundary from
%	its cumulants?? } \uv{Giving and discussing asymptotic is fair and done below. Yet I am afraid it's not a great idea to re-summon the fictitious beasts that the whole foregoing text advocated against. }
% \red{can you make this more precise?}.}
To interpret the divergent part, let us recall that when $(a_1,a_2)$ crosses one of the boundaries of $\mathcal{R}$, the 2D logREM goes through a phase transition~\cite{carpentier2001glass,fyodorov2009statistical,cao16liouville,cao17seiberg}.  Nicely, the divergences of the variance corrections $C_{2}$ can be interpreted as the integrable signature of phase transitions. Those of the other cumulant corrections can be also rationalized, albeit on a more formal level. Since there are two types of them, we discuss separately below [see Fig.~\ref{fig:C2D} (a)]:
\begin{itemize}
	 \item[-]  When $a_1 + a_2$ decreases below $1$,  the 2D logREM goes through a long-distance \textit{un}binding (or escaping~\cite{cao16liouville}) transition. The potential $U(z)$ can no longer confine the thermal particle in an $O(1)$-size region around $0$ and $1$. Then, the free energy/minimum of the 2D logREM behaves as in an ordinary logREM without background potential. In particular, its variance is of order unity, parametrically  (in $L$)  smaller than  compared to eq.~\eqref{eq:2Dcorrections}:
	 \begin{equation}
	 \left. \overline{V_{\min}^2}^c\right\vert_{a_1 + a_2 < 1} \sim O(1) \ll 2\ln (L^2) + C_2 =
	 \left. \overline{V_{\min}^2}^c\right\vert_{a_1, a_2 \in \mathcal{R}}   \,,\, L \to \infty \,. \label{eq:escapingvar}
	 \end{equation}
	 Now, at the brink of the binding transition, i.e., as $a_1+a_2$ approaches $1_-$, the minimum variance correction $C_2 \to -\infty$: such a negative divergence is the precursor signature of the parametric decrease caused by the long-distance escaping transition. Since $C_2$ is not a variance itself but a correction thereof, its negativity is not problematic.
	
   In general, $C_k$ diverges negatively for all $k$, see eq.~\eqref{eq:Cksingular} and Fig.~\ref{fig:C2D} (c). Formally this is due the fact that when $a_1 + a_2 \searrow Q/2$ ($Q=2$ in the low-temperature phase), the zero of the moment generating function $\overline{\exp(t V_{\min}))}$ at $t = t_* =  a_1 + a_2 - Q/2 \nearrow 0$ approaches from the right, whose physical significance is discussed in section \ref{sec:DF}. Such a zero contributes a term $\sim \ln (1 - t / t_*)$ to the cumulant generating function, thus a negatively divergent contribution $-t_*^{-k}(k-1)! \to -\infty$ to the $k$-th cumulant, consistent with eq.~\eqref{eq:Cksingular}.
	
	 	\item[-]  When $a_1$ increases beyond $1$, the 2D logREM goes through a \textit{short-distance binding transition}: its Gibbs measure becomes concentrated in an $\epsilon$-size region around the associated log-singularity $z = 0$ of the deterministic potential $U(z)$, eq.~\eqref{eq:Uz}; we recall that $\epsilon$ is the short-distance cut-off of the 2D logREM, see eq.~\eqref{eq:ZDF}~\footnote{the discussion applies to $a_2$ in lieu of $a_1$ by symmetry.} . In this short-distance bound phase, the minimum variance will become parametrically (in $1/\epsilon$) larger compared to eq.~\eqref{eq:2Dcorrections} (see e.g., \cite{cao17thesis}, section 2.3.4):
	 \begin{equation} \left. \overline{V_{\min}^2}^c\right\vert_{a_1>1} \sim 2\ln (L^2/\epsilon^2) \gg 2\ln (L^2) + C_2 =
	 \left. \overline{V_{\min}^2}^c\right\vert_{a_1, a_2 \in \mathcal{R}} \,,\, \epsilon\to 0 \,. \label{eq:bindingvar}
	 \end{equation}
	 Now, at the brink of the binding transition, i.e., as $a_1$ or $a_2$ approaches $1_-$, the minimum variance correction $C_2 \to +\infty$: such a positive divergence is the precursor signature of the parametric increase caused by the phase transition.
	
	  In general, $C_k$ diverges positively (negatively) if $k$ is even (odd, respectively), see eq.~\eqref{eq:Cksingular} and Fig.~\ref{fig:C2D} (c). Formally, this is due the fact that when $a_1 \nearrow Q/2$  ($Q=2$ in the low-temperature phase), a negative pole of the moment generating function $\overline{\exp(t V_{\min})}$ at $t = t_* = 2a_1-Q \nearrow 0$ approaches $0$ from the left. This pole comes from the factor $\Barnes_{b}(Q-a_1 + t)$ in eq.~\eqref{eq:continuedDF}, see also eq.~\eqref{eq:Barneszeros}.  Such a pole appears generally in logREMs with a charge  $a_1 < Q/2$~\cite{fyodorov2009statistical}, and is related to a universal negative exponential tail of form $e^{-\mathcal{F}_{2D} t_*}$ of the free energy distribution~\cite{cao17seiberg} (Fig 3 b and Appendix B). Such an exponential distribution has $t_*^{-k}(k-1)!$ as $k$-th cumulant, coinciding with the corresponding divergent part of eq.~\eqref{eq:Cksingular}. Therefore, in this case, the divergent part has a stand-alone statistical interpretation.
\end{itemize}
 }

 \bibliography{rems}

\end{document}